\documentclass[graybox,spbasic]{svmult}

\usepackage{amsmath}
\usepackage{mathptmx}       
\usepackage{helvet}         
\usepackage{courier}        
\usepackage{type1cm}        
\usepackage{makeidx}         
\usepackage{graphicx}        
\usepackage{multicol}        
\usepackage[bottom]{footmisc}

\makeindex             
\usepackage{aas_macros}        

\newcommand{\simgreat} {\mathbin{\lower 3pt\hbox{$\rlap{\raise
        5pt\hbox{$\char'076$}}\mathchar"7218$}}}

\newcommand{\simless}{\mathbin{\lower 3pt\hbox {$\rlap{\raise
        5pt\hbox{$\char'074$}}\mathchar"7218$}}}

\begin{document}

\title*{Polarimetry of binary systems: polars, magnetic CVs, XRBs}
\titlerunning{Polarimetry of binary systems} 

\author{Tariq Shahbaz}

\institute{Tariq Shahbaz \at 
Instituto de Astrof\'\i{}sica de Canarias (IAC), E-38200 La Laguna, 
Tenerife, Spain and
Departamento de  Astrof\'\i{}sica, Universidad de La Laguna (ULL), 
E-38206 La Laguna, Tenerife, Spain, 
\email{tsh@iac.es}
}
%
%
\maketitle

\abstract { 
Polarimetry provides key physical information on the properties of 
interacting binary systems, sometimes difficult to obtain by any other type 
of observation. Indeed, radiation processes such as scattering by free 
electrons in the hot plasma above accretion discs, cyclotron emission by 
mildly relativistic electrons in the accretion shocks on the surface of 
highly magnetic white dwarfs and the optically thin synchrotron emission 
from jets can be observed. In this review, I will illustrate how optical/near-infrared polarimetry 
allows one to estimate magnetic field strengths and map the accretion zones 
in magnetic Cataclysmic Variables as well as determine the location and 
nature of jets and ejection events in X-ray binaries.
}

\section{Cataclysmic variables }

\begin{figure}[]
\sidecaption
  \includegraphics[scale=0.4]{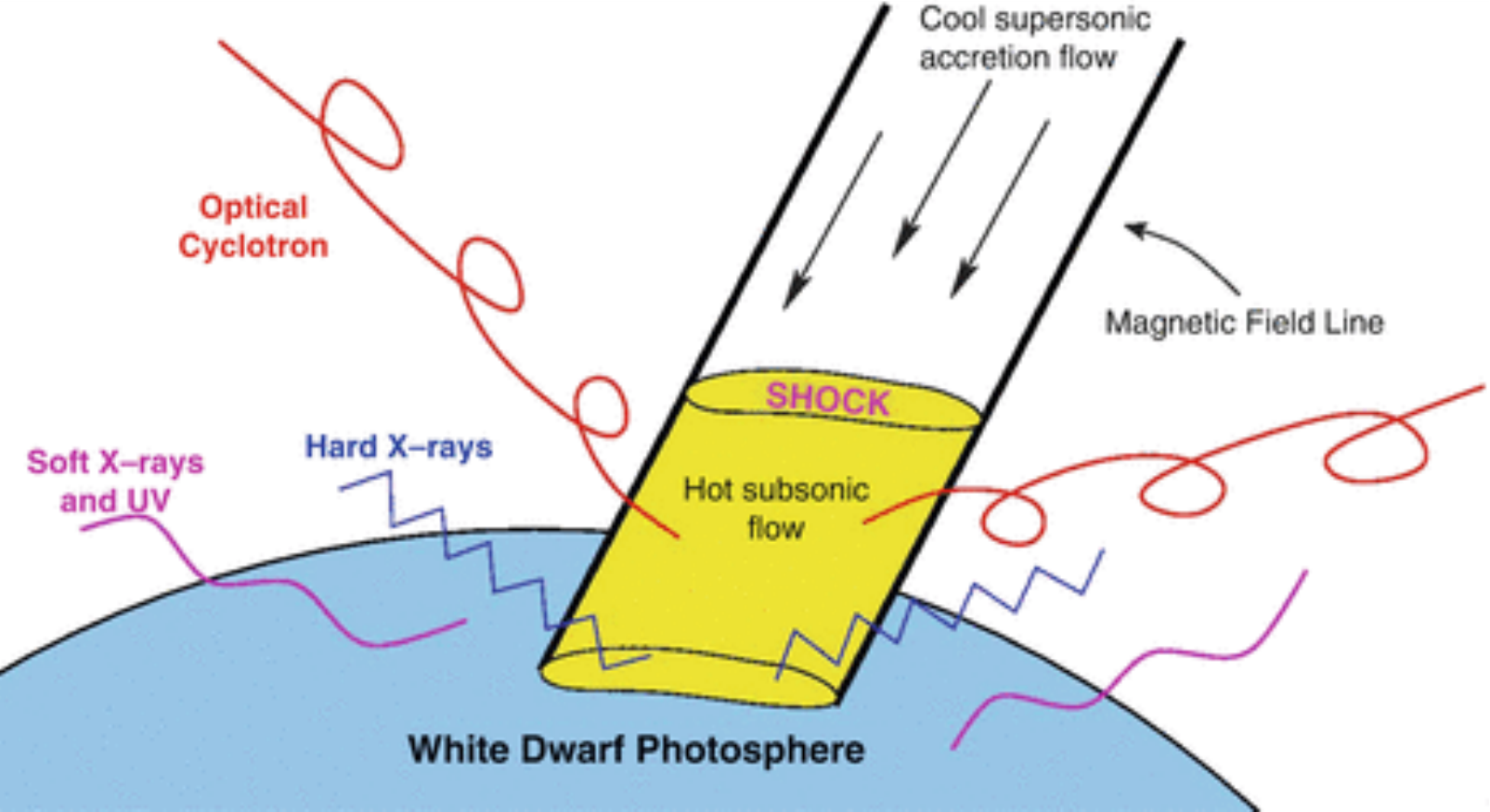}
\caption{
A schematic of the surface of the WD, showing the region of the main emission
mechanisms \cite{Potter16}.
}
\label{fig:schem_accn}       
\end{figure}

Cataclysmic variables (CVs) are a class of binary system in which a main 
sequence secondary star transfers matter to its white dwarf (WD) companion, 
via Roche lobe overflow. It is estimated that $\sim$20\% of all CVs host a 
magnetic WD (mCVs) \cite{Wickramasinghe00, Ferrario15}. In the 
"normal" accreting mode known as the "high-state" of mCVs, the accreting 
material leaving the inner Lagrangian L1 point attaches itself to the 
magnetic field lines of the WD close and is channelled towards its 
surface. However, most mCVs also show erratic and abrupt "low-states," 
during which accretion through the L1 point temporarily stops, presumably 
due to changes in the mass-transfer rate in the system, related to the 
atmospheric magnetic activity of the secondary star \cite{Livio94}. During 
the high-state, as the flow is collimated by the magnetic field, electrons 
spiral down the magnetic field lines, cooling the plasma flow by cyclotron 
emission. This radiation is detected as broad overlapping cyclotron humps 
in the optical and infrared (IR) spectrum and allows to estimate 
magnetic field strengths within the accretion region. Hard X-rays resulting from a 
post-shock region just above the accreting pole and softer ultraviolet (UV) 
and X-rays 
are emitted from a hot photospheric polar cap. During the low-states, 
Zeeman spectroscopy can be used to constrain the temperature and magnetic 
field of the WD (see Fig. \ref{fig:schem_accn}; for a review see 
\cite{Cropper90}).

\begin{figure}[t]
\begin{center}
  \includegraphics[scale=0.4]{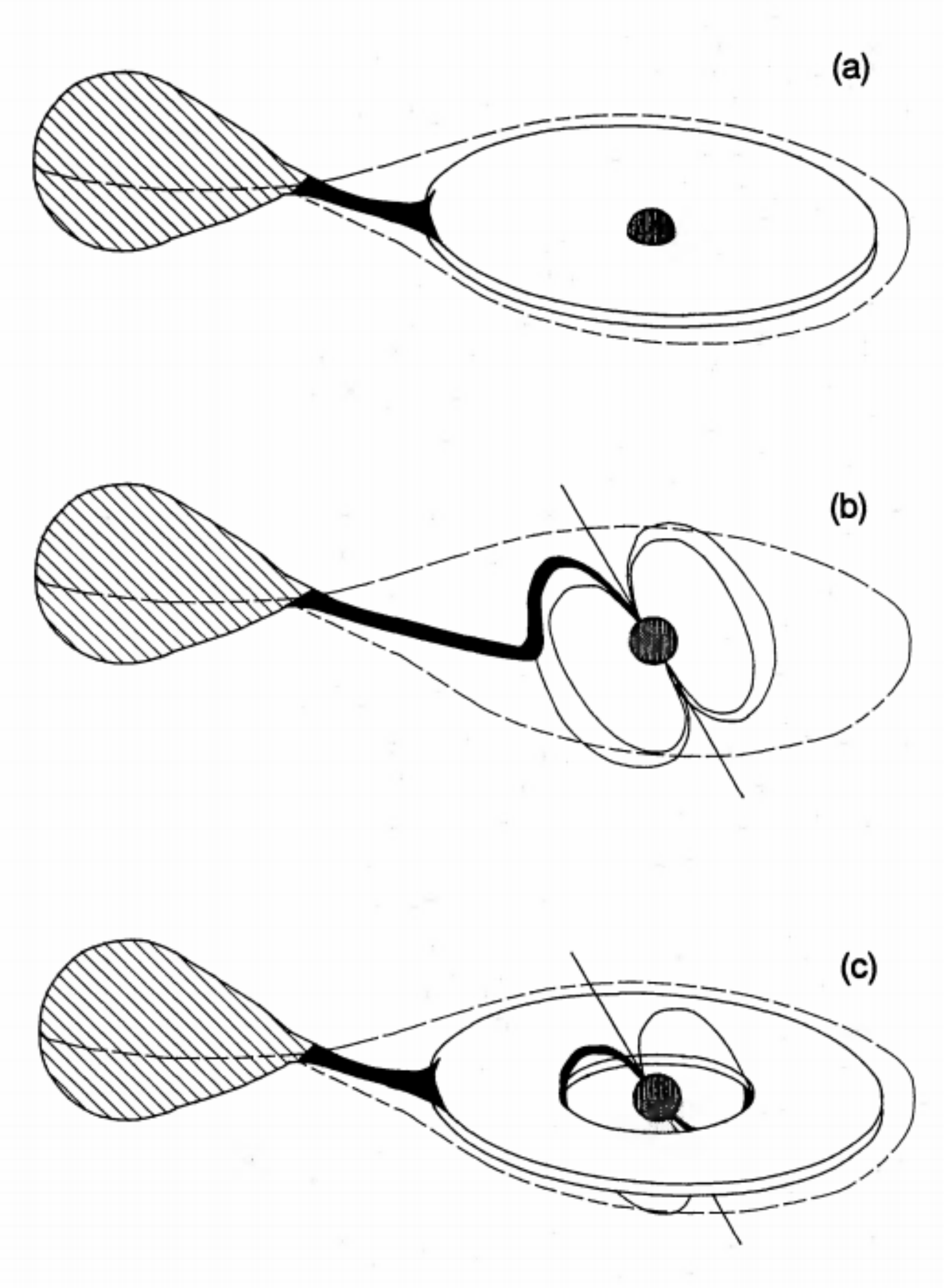}
\end{center}
\caption{
Schematic of the different type of CVs. From top to bottom; a non-magnetic CV
with an accretion disc (a),   a polar (AM Her type systems) in which the
accretion disc is prevented from forming  (b) and an intermediate polar (DQ Her
type systems) in which the accretion disc is disrupted  (c) \cite{Norton93}.  
}
\label{fig:schem_cvs}       
\end{figure}

\noindent
MCVs are characterized by strong X-ray emission, high-excitation optical 
spectra, very stable X-ray and optical periods in their light curves. They 
are broadly classified into two categories based upon their periodicities, 
which are in turn assumed to imply different magnetic field strengths. The 
polars (or AM Her type systems after the prototype system)  contain 
synchronously rotating WDs, whilst in intermediate polars (IPs or DQ Her 
type systems) the WDs rotate asynchronously with respect to the binary 
orbit (see Fig. \ref{fig:schem_cvs}).

\begin{itemize}

\item
In polars, the magnetic field strength of the WD ($>$10 MG) is so 
strong that the gas leaving the inner Lagrangian point L1 following a 
ballistic trajectory, eventually attaches to the magnetic field of the WD, 
preventing the formation of an accretion disc as in CVs. Accretion can 
occur on one or both poles, depending on the orientation of the magnetic 
dipole of the WD.  The material is fully ionized producing highly polarized 
cyclotron emission due to the presence of the strong magnetic field. Since a 
large fraction of the optical flux in polars comes from such regions, they 
have large polarization values. The radiation is circularly polarised 
when observed parallel to the magnetic field lines and linearly when 
observed perpendicular to the field lines. Consequently, polarimetry is a 
fundamental tool for classifying an object a polar. The magnetic field of 
the WD extends to the secondary synchronizes the spin and orbital period. 
Typically the polars are found to have an orbital and spin period of less 
than two hours (for a review see \cite{Cropper90}).

\item
The IPs are asynchronous systems which have a magnetic field 
strength of $<$10 MG. The lower magnetic field strength compared to 
polars implies that matter falls inward towards the WD in much the same way 
as in non-magnetic CVs. However, the magnetosphere disrupts the inner 
portion of the accretion disc and the flow then follows the field lines in 
a curtain-like structure \cite{Patterson94}. Due to the disc symmetry, the 
accretion curtains are present for both magnetic poles, resulting in 
arc-like accretion regions near the WD  poles, similar to the 
auroral ovals on the Earth and Jupiter.
\end{itemize}

\noindent 
However, there are systems that do not fit into the standard picture of mCVs.
AE\,Aqr was originally classified as an IP \cite{Patterson79} but there are a 
number of unusual observational  features that are not naturally explained by
this model It is known to switch between the flaring and quiescent states
irregularly, which distinguishes it from other mCVs.  Studies of the 33 s WD
pulsations in the optical/UV \cite{Eracleous94} and X-rays \cite{Reinsch95,
Choi99} show no correlation between their amplitudes and the flaring of the
system, which means that  the flares are not related to the process of
depositing material onto the WD. \cite{Eracleous94}. The Balmer emission lines
are single peaked and produce Doppler tomograms that are not consistent with
those of an accretion disc.  \cite{Welsh98}. Different models were proposed to
explain the observational properties of AE Aqr.  The first, a "magnetic
propeller'' model \cite{Wynn97} suggests that the rotation rate of the WD
decelerates by means of interaction between its fast rotating magnetosphere and
the material inflowing from the secondary.  The second "pulsar-like white
dwarf'' model \cite{Ikhsanov98} show that  the observed braking of the white WD
can be explained in terms of the canonical pulsar-like spin-down mechanism
\cite{Ikhsanov98, Ikhsanov06}. Recently, a numerical  propeller model of AE\,Aqr
using  axisymmetric magneto-hydrodynamic (MHD) simulations has been developed
\cite{Blinova18}. They suggest that an accretion disc  forms around the WD that
interacts with the magnetosphere of the star in the propeller regime. 
The model explains
the rapid spin-down of the WD through the outflow of angular momentum from the
surface of the WD into the wind and 
predicts  that $\sim$85\% of the
inner disk matter is ejected into the conically-shaped winds. 
The model can also explain the low accretion rate onto
the star if the radiative efficiency of accretion is $\sim$7\%,

\subsection{Models}

The optical flux and polarization of polars is highly modulated with 
orbital phase. These observations are explained by the cooling of the 
post-shock region through cyclotron emission which depends strongly on the 
viewing angle and the physical properties of the emitting region. Stokes 
techniques are consequently a powerful tool to diagnose the physical properties of the accreting region. An increase in circular 
polarization is observed when one looks down on the shock (parallel to the 
magnetic field lines), whereas an increase in linear polarised flux is 
observed when one views the side of the shock (perpendicular to the 
magnetic field lines).

\begin{figure}[b]
\begin{center}
  \includegraphics[scale=0.7]{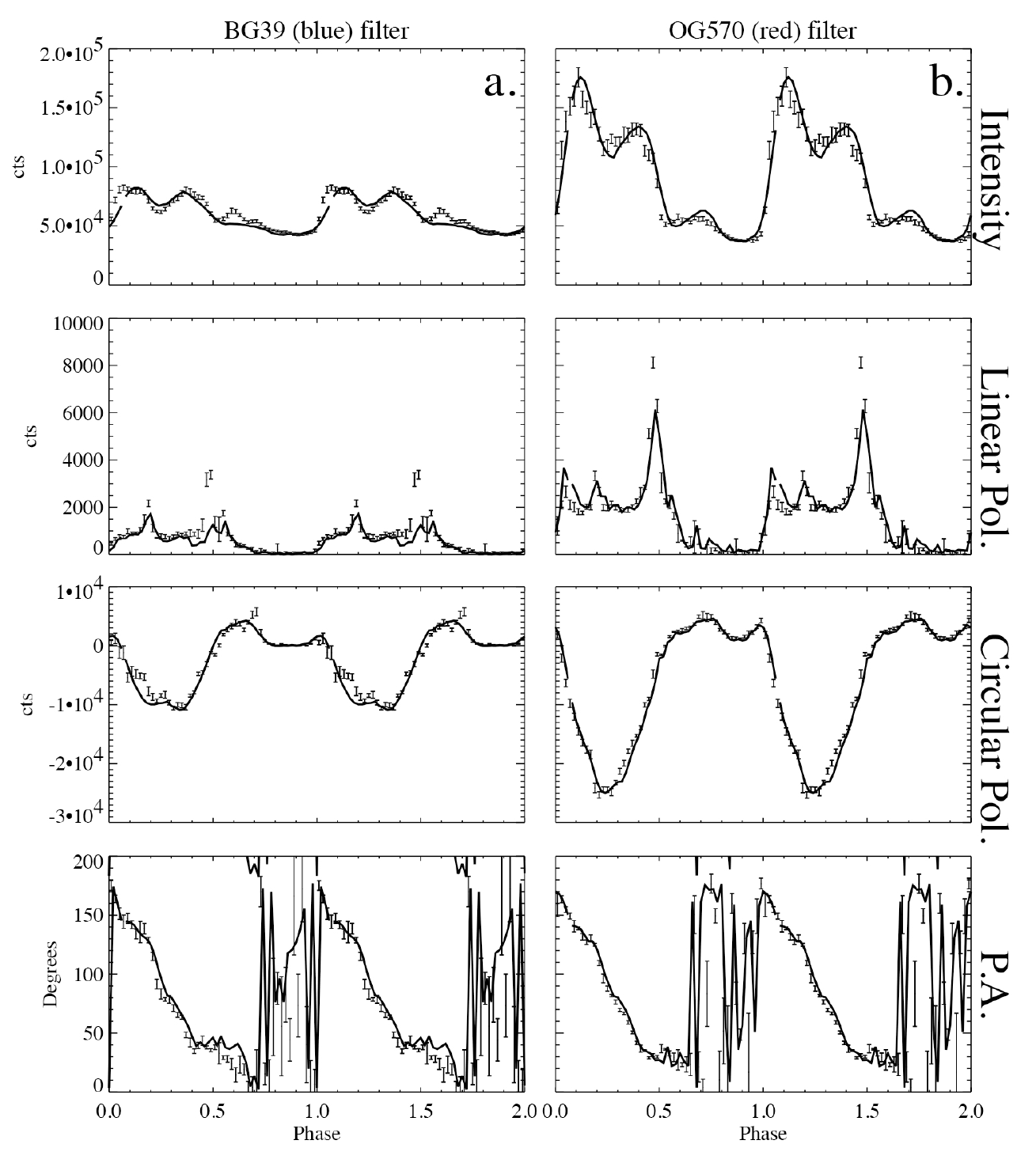}
\end{center}
\caption{
The Stokes imaging optimized solution to the polarimetric observations 
V347\,Pav. Solid curves are the model fit to the blue (a) the red (b) data. 
The imaging predicts that the 
bright-phase emission arisws from a large region towards the lower 
hemisphere, and the faint-phase emission from a single smaller region in 
the upper hemisphere of the WD \cite{Potter00}.
}
\label{fig:stokes}       
\end{figure}

The first important step in modelling of polar observations were performed 
in the 1980's and assumed the emitting region to be a 1-D point source. 
\cite{Chanmugam81, Meggitt82, Wickramasinghe85}. Wickramasinghe \& Meggitt 
(1985) present a grid of fluxes, and linear and circular polarization as a 
function of viewing angle and harmonic, which can be compared to optical 
data of polars. However, there is observational evidence of extended 
emitting regions in polars and so the models were improved to 2-D by using 
a sum of distinct point sources to represent the emitting region 
\cite{Ferrario90, Potter98}.

More sophisticated models and fitting techniques were then developed 
referred to ''Stokes imaging'' \cite{Potter98} was the first objective and 
analytical technique that robustly models the cyclotron emission and maps 
the accretion zones in magnetic mCVs. This technique allowed objective 
mapping of the cyclotron-emission regions in terms of location, size and 
density structure. Stokes imaging uses a cyclotron-emission model and a 
grid of cyclotron-emission calculations which include cyclotron and 
free-free opacities \cite{Wickramasinghe85}. The emission from a single 
emission spot on the surface of the white dwarf is modelled by first 
calculating the local magnetic field strength and subsequently the local 
optical depth parameter. The cyclotron emission grids are then used to 
produce variations in the polarized emission as a function of phase. The 
emission from extended sources is modelled by summing the components of 
many such emission spots. Projection effects of extended regions are also 
taken into consideration. Fig. \ref{fig:stokes} shows the model fit from 
Stokes imaging for the IP V347\,Pav, where Stokes imaging reproduced the 
main polarization morphology \cite{Potter00}. The main emission region was 
found to consist with two relatively higher density regions centred on the 
magnetic equator. The system inclination derived is 68 degrees and the 
magnetic field strengths are 15 and 20 MG for the upper and lower poles 
respectively. Stokes imaging has been advanced by the addition of more 
realistic stratified accretion shock model in order to calculate cyclotron 
spectra, and the use of Stokes imaging in combination with Doppler and 
\cite{Marsh88} Roche tomographic \cite{Watson01} techniques, to gain 
insights into magnetic accretion \cite{Potter04}.

\begin{figure}
\begin{center}
  \includegraphics[scale=0.38]{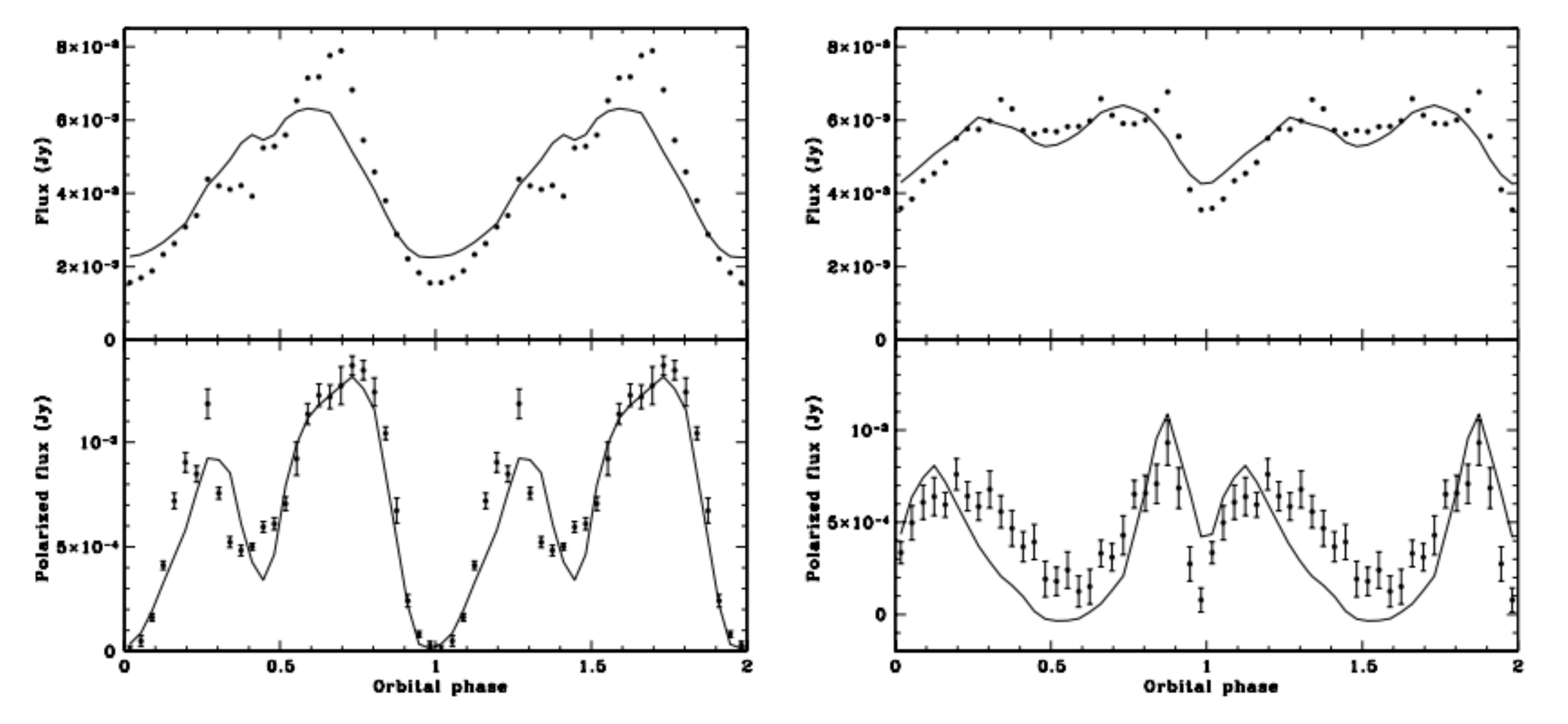}
\end{center}
\caption{
Best fit to the $V$-band (left) and $J$-band (right) circular polarization 
light curves V834\,Cen of using the \textsc{cyclops} \cite{Costa09}. 
}
\label{fig:cyclops}       
\end{figure}

A more detailed code has also been developed, which adopts a 3-D 
representation of the accretion column and WD, incorporates the radiative 
transfer of the cyclotron process in the post-shock region \cite{Costa09}. 
The model has been used to fit the optical near-IR circular polarimetric 
light curves of V834\,Cen (see Fig. \ref{fig:cyclops}) and the best fit is consistent with previous 
parameter estimates and is able to reproduce the observed features.

\subsection{Cyclotron spectroscopy}

Cyclotron spectroscopy is a useful method for extracting on strength of the 
magnetic fields in polar and IPs. The characteristic feature of polars is 
the presence of strong linear and circular polarization at optical and 
near-IR wavelengths. The polarized emission is due to cyclotron origin and 
arises from accretion shocks near the surface of the high magnetic field 
WD. The fundamental cyclotron frequency usually lies in the IR (from 
$\sim$2--10 $\mu$m). Indeed, the cyclotron radiation observed in the 
optical/near-IR consists of the harmonics of the fundamental cyclotron 
frequency.  The spacing between successive cyclotron humps allows one to 
estimate the magnetic field strength of the cyclotron emission, because 
cyclotron emission intensity and polarization depend on the harmonic $n$ 
(=$\omega/\omega_c$) of the fundamental cyclotron frequency $\omega_c$ 
\cite{Wickramasinghe85, Ferrario93}. The position of the $n_{th}$ harmonic 
$\lambda_n$ of the cyclotron emission from a region with magnetic field $B$ 
and low electron temperatures ($T_e<$10 keV), as a first approximation 
is given by

\begin{equation} 
\lambda_n \sim \frac{10,710}{n}  \left( \frac{10^8}{B}\right) \, \AA
\end{equation}

\noindent
Cyclotron line emission has been observed in a number of polars, where
cyclotron harmonics ($n\sim$3 to 8) are seen in the optical/near-IR spectra.
W\,Pup, \cite{Wickramasinghe89},  UZ\,For \cite{Ferrario89},  MR\,Ser
\cite{Wickramasinghe91},  DP\,Leo \cite{Cropper93} ST\,LMi \cite{Ferrario93}
and  EF\,Eri \cite{Campbell08}.

\begin{figure}[t]
\begin{center}
  \includegraphics[scale=0.45]{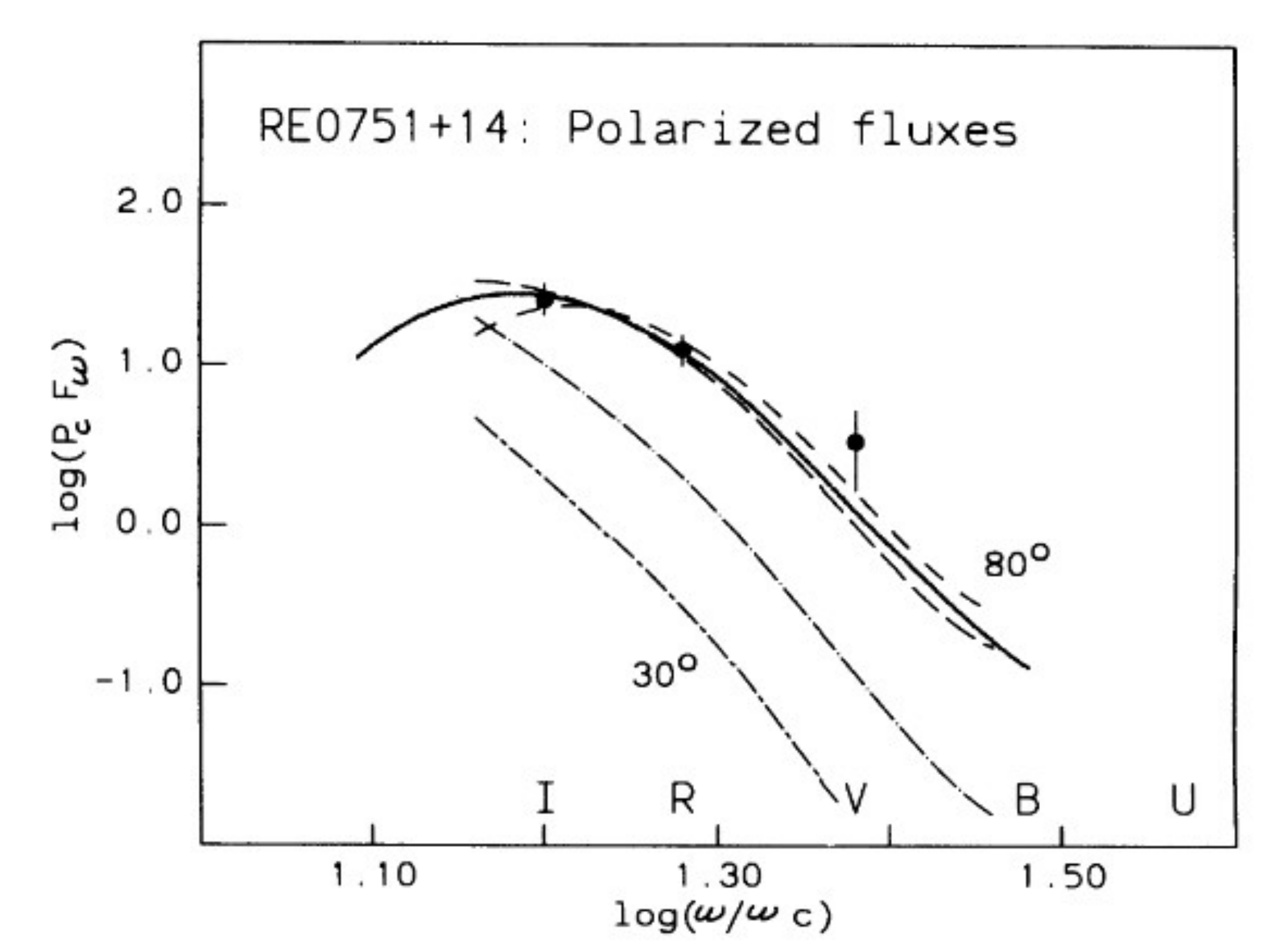}
\end{center}
\caption{
Observed peak circular polarized flux compared with cyclotron models with 
different magnetic field strengths \cite{Piirola93}.
\copyright AAS. Reproduced with permission.
}
\label{fig:polspec}       
\end{figure}

The observed circular polarized flux is independent of other sources of 
non-polarized light (e.g. accretion stream, disc or WD photosphere) and 
therefore is very useful for comparing with existing cyclotron models.  
The broadband circular polarized flux spectrum gives estimates to the 
magnetic field in the cyclotron emission region, as cyclotron emission 
intensity and polarization are characteristic to each harmonic 
\cite{Wickramasinghe85}.  An example of the comparison of the observed 
circular polarized flux spectrum with cyclotron models is shown in Fig. 
\ref{fig:polspec}. IPs are known to be circularly polarised at a level 
$\sim$1--3\% \cite{Ferrario15}.  Magnetic field measurements for IPs which 
do not show measurable photospheric Zeeman features are difficult, but from 
the polarized flux spectrum, estimates have been made for 
PG\,Gem (8--18 MG \cite{Piirola93}); 
PQ\,Gem (9--21 MG \cite{Vaeth96, Potter97}), 
V2400\,Oph (9--20 MG \cite{Vaeth97}), 
RX\,J2133.7+5107 ($>$20 MG \cite{Katajainen07}) and 
V405\,Aur ($\sim$30 MG \cite{Piirola08}). These values are similar
to the magnetic field strengths measured  in polars, and suggest that some of 
the polarized IPs may evolve into polars \cite{Chanmugam84}.

\subsection{Time-resolved polarimetry of SW\,Sex systems}

SW\,Sex stars are a sub-class of CVs that possess common but inexplicable 
features, very different from other CVs \cite{Thorstensen91}. Their light curves show deep eclipses suggesting a high-inclination 
accretion disc but their emission lines show single-peaked profiles that do 
not share the orbital motion of the WD and they  exhibit transient absorption 
features as well as shallow eclipses. This is contrary to the 
deeply-eclipsing, double-peaked profiles one would expect for 
high-inclination systems \cite{Marsh88}. Many  models 
have been proposed to explain the behaviour of the SW\,Sex systems. Models 
include gas-stream overflow, magnetic propeller accretion, and self 
obscuration by the disc rim (for details see \cite{Rodriguez07}). Dhillon et 
al. (2013) have proposed that a combination of dominant bright-spot 
emission and a self-occulting disc can explain the observed properties.

\begin{figure}[t]
\begin{center}
  \includegraphics[scale=0.3]{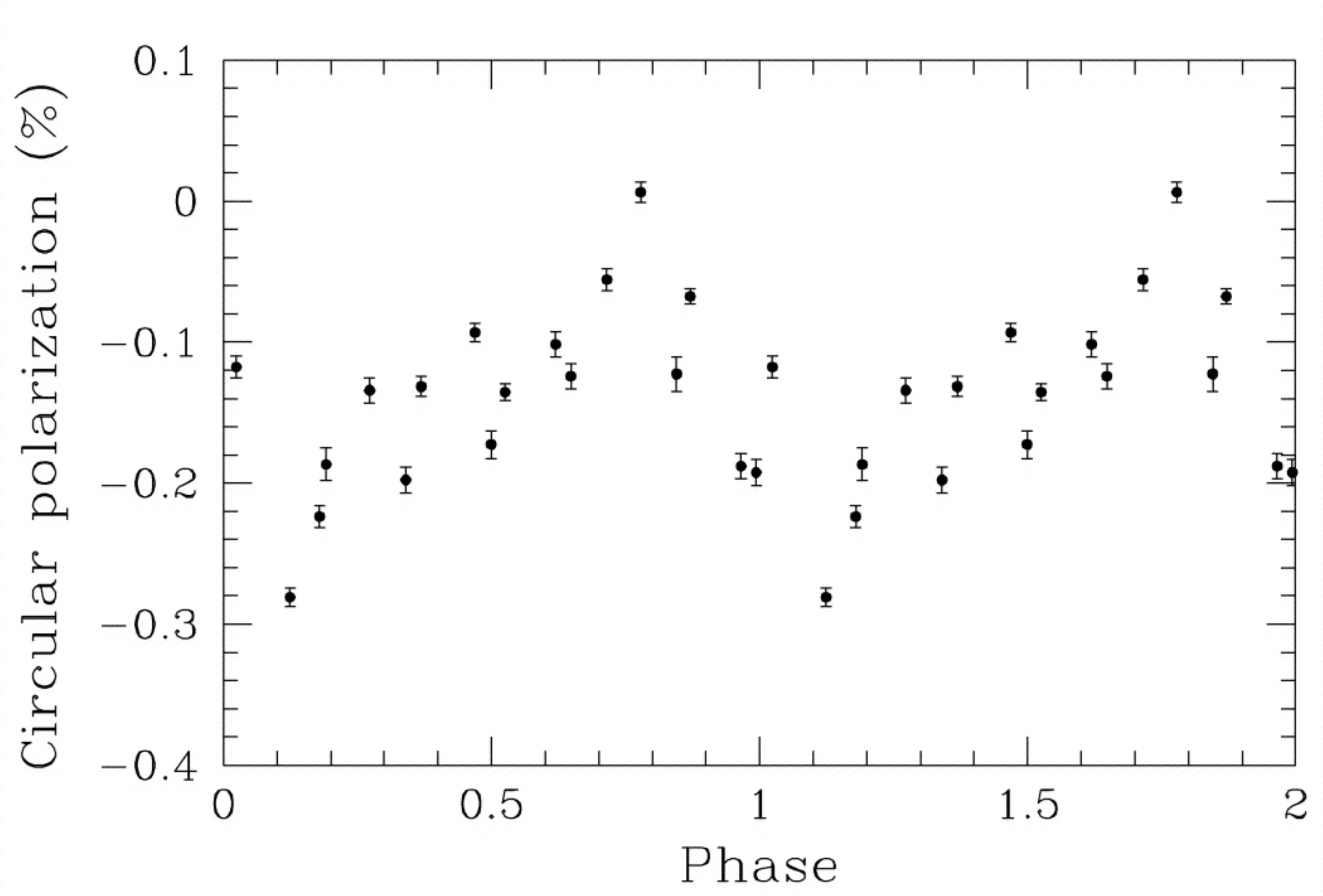}
\end{center}
\caption{ 
Modulated circular optical polarization in  LS\,Peg. The period of the modulation is in phase with the spin period of the WD  \cite{Rodriguez01}.
\copyright AAS. Reproduced with permission.
}
\label{fig:lspeg}       
\end{figure}

A magnetic origin for the SW\,Sex systems has been claimed by several 
authors (e.g. see \cite{Martinez99}). Although the observation of modulated 
circular polarisation is an unambiguous proof as it implies the direct 
detection of a relatively strong magnetic field, the presence of coherent 
pulsations related to the WD spin period and/or the orbital period also 
provides strong evidence for the presence of a magnetic WD.  Modulated 
circular polarization has been observed in the non-eclipsing systems 
LS\,Peg (see Fig. \ref{fig:lspeg}), V795\,Her and RXJ1643.7+3402 
\cite{Rodriguez01,Rodriguez02,Rodriguez09}. In LS\,Peg X-ray pulsations 
associated with the spin period of the WD is observed at the same 
polarimetric frequency \cite{Baskill05}, whereas in RXJ1643.7+3402 the 
circular polarization period is interpreted as half the beat period between 
the WD spin period and the orbital period \cite{Rodriguez09}. To some 
extent the observed circular polarization frequencies can be explained in 
terms of a magnetic model in which the overflowing gas stream crosses the 
WD's magnetosphere at approximately the corotation radius 
\cite{Rodriguez01, Rodriguez09}.

\subsection{A WD pulsar}

The discovery of pulsar-like behaviour in the close binary system AR 
Scorpii (hereafter AR\,Sco)  demonstrates that WDs can exhibit 
many of the same characteristics as neutron star (NS) pulsars. Highly 
pulsed non-thermal emission is observed from radio to the UV arising 
from the 
spinning WD \cite{Marsh16}.

These pulsations are seen predominantly at the 
118 s beat period between the 117 s spin period and the 3.6 h orbital 
period of the binary. The strong spin modulated emission arising from the 
WD plus the dominant non-thermal nature of the spectral energy 
distribution suggests that magnetic interactions 
power the emission in AR Sco \cite{Marsh16, Katz17}, similar to what is 
observed in pulsars. Further evidence of pulsar behaviour is given 
by polarimetric observations \cite{Buckley17}. Strong linear polarization 
(up to 40\%) that varies on both the spin period of the WD and the beat 
period between the spin and orbital period has been detected as well as 
low-level circular polarization ($\sim$ few \%).
The characteristics 
of the polarized and non-polarized emission are explained in terms of 
synchrotron emission from two different regions, one associated with the 
rotating WD magnetic field and the other as a result of MHD interactions 
with the M-dwarf companion star. Unlike typical CVs there appears to be 
little accretion on to the WD from its M-dwarf companion star 
\cite{Marsh16}, which means that gravitational potenitial energy is not 
the source of the emission. AR Sco is the first example of a WD pulsar. 
because the pulsed luminosity of AR\,Sco is powered by the spin-down of 
the rapidly rotating highly magnetized ($<$500 MG) WD \cite{Buckley17}.

\begin{figure}[t]
\begin{center}
  \includegraphics[scale=0.55]{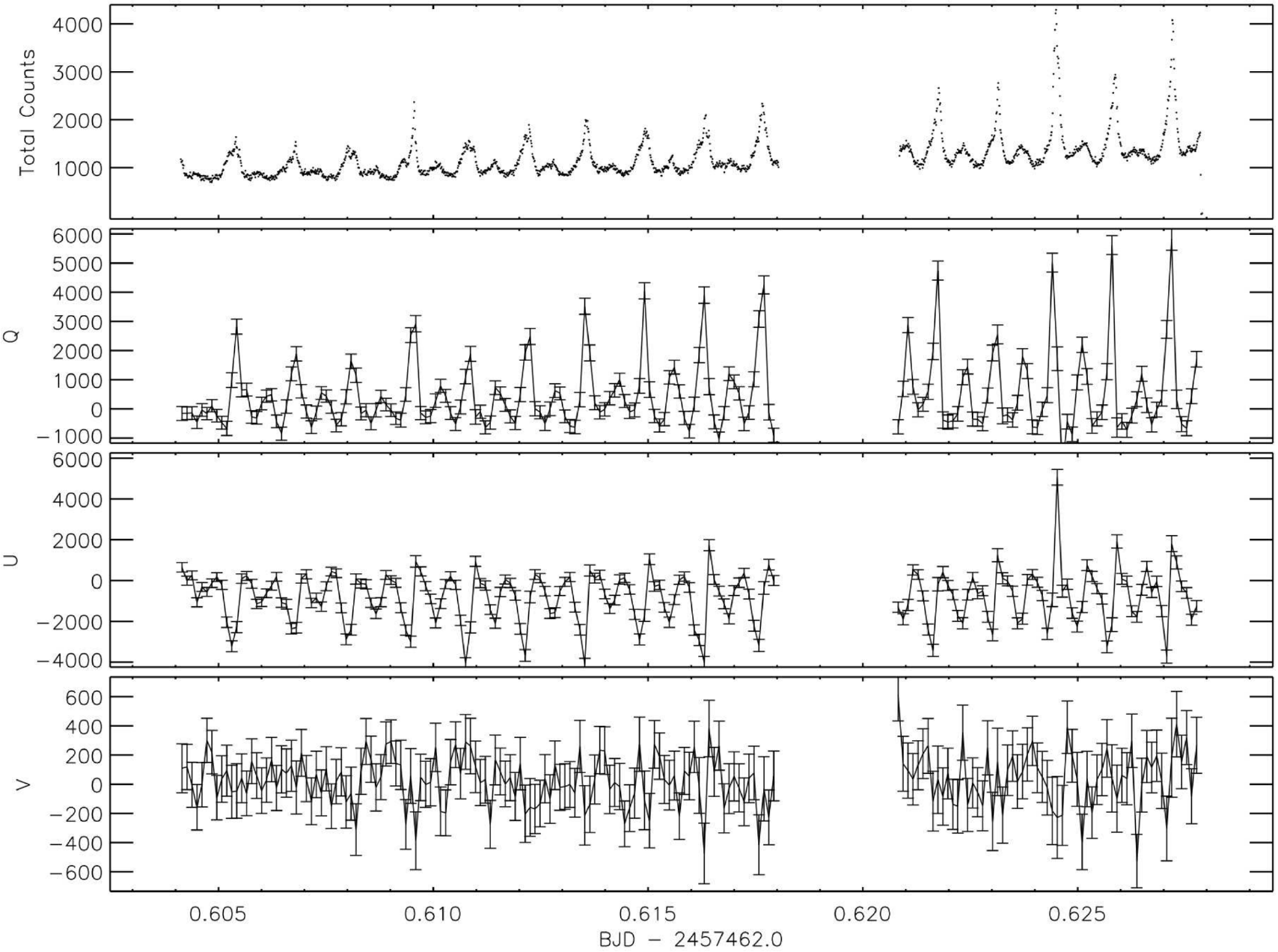}
\end{center}
\caption{
Broad band photopolarimetry of AR\,Sco. From the top to bottom, 
the total intensity and the Stokes Q, U \& V values \cite{Buckley17}.
}
\label{fig:sed}       
\end{figure}

\begin{figure}[]
\begin{center}
  \includegraphics[scale=0.55]{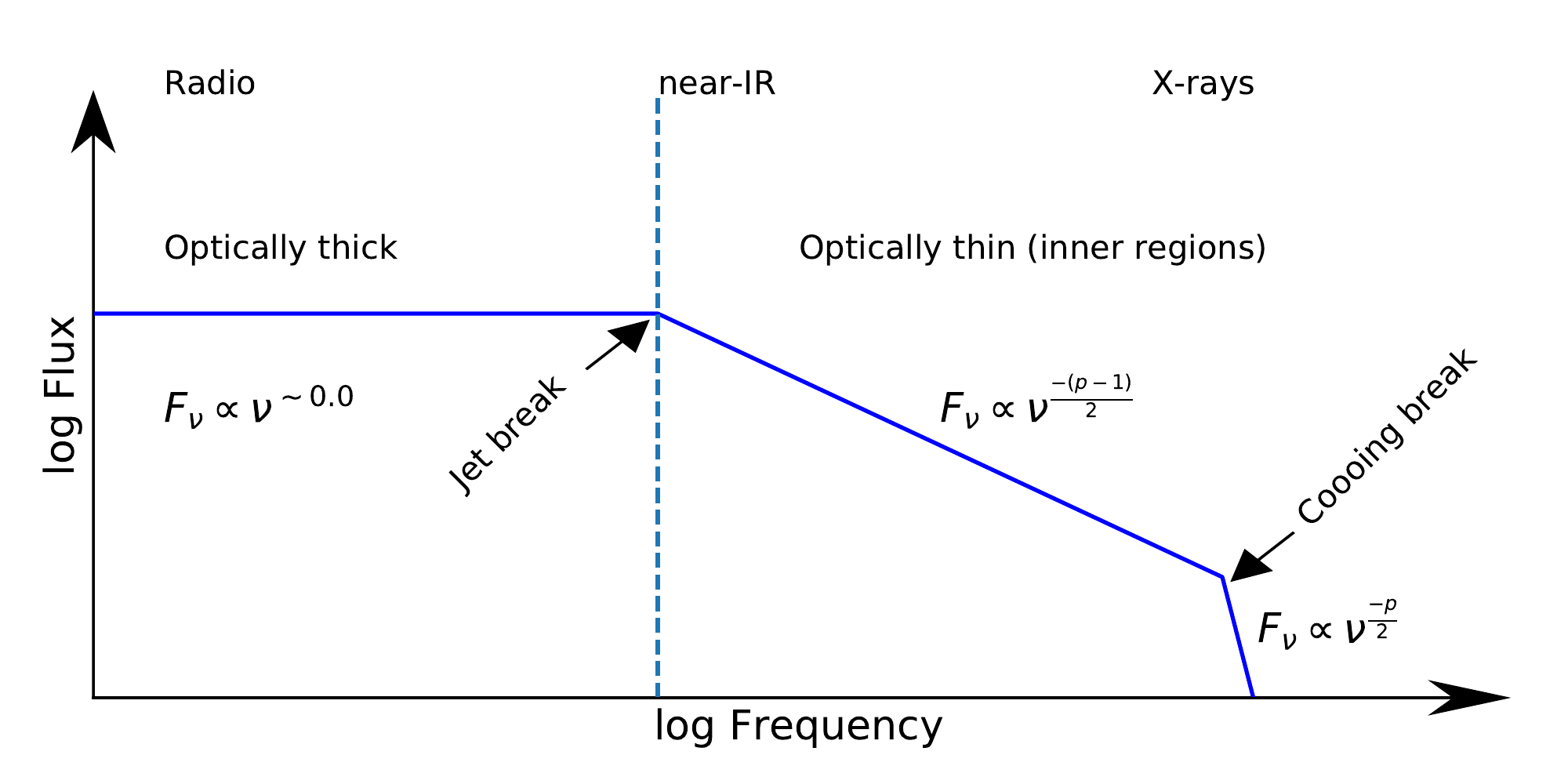}
\end{center}
\caption{
A sketch of spectrum of a steady, hard-state jet. The radio to near-IR spectrum 
is due to a $\sim$ flat, self-absorbed optically thick synchrotron spectrum.
The key spectral point is the break from optically to optically thin
synchrotron emission, between the mid- to near-IR wavelengths. 
}
\label{fig:sed}       
\end{figure}

\section{X-ray binaries}

X-ray binaries (XRBs) are binary systems in which a compact object, a black 
hole (BH) or a NS, accretes matter from a companion star.  
During the outburst of a black BH-XRBs, continuously replenished 
synchrotron-emitting "compact" jets are commonly produced during the hard 
spectral state (hard-state; \cite{Fender14}). (For descriptions of X-ray states see 
\cite{MR06}).  The hard-state usually occurs at the beginning and end of an 
outburst, with X-ray luminosities $\rm >10^{33}\,\,erg\,s^{-1}$ and up to 
$\sim$1--5\% of the Eddington luminosity (e.g. \cite{Belloni10, 
Koljonen16}), with a highly variable hard power-law X-ray spectrum that 
originates from the inner, radiatively inefficient accretion flow close to 
the black BH \cite{Narayan94}. A classical "flat spectrum" radio jets 
(similar to those seen in AGN) are commonly observed during the hard-state, 
when the accretion flow structure allows a large-scale height magnetic 
field. In the softer X-ray states jets are dramatically quenched at radio 
frequencies \cite{Fender01, Gallo03, Russell11a} because the geometrically 
thin disc suppresses any large-scale vertical field \cite{Meier01}.

Few studies have attempted to uncover the polarimetric signature of the 
optically thin synchrotron emission from compact jets in XRBs.  The 
polarization originates close to the base of the jet where the magnetic 
field should have higher level of ordering over the smaller emission region compared to 
further out in the jet \cite{Blandford79}.  
Polarimetric measurements of the optically thin power law synchrotron 
emission provides a powerful tool to uncover the nature of the magnetic 
field structure in this region, because it is associated with the start of the 
particle acceleration in the jet \cite{Polko10}, which is important for 
models and simulations of jet production.

\begin{figure}[t]
\begin{center}
  \includegraphics[scale=0.35]{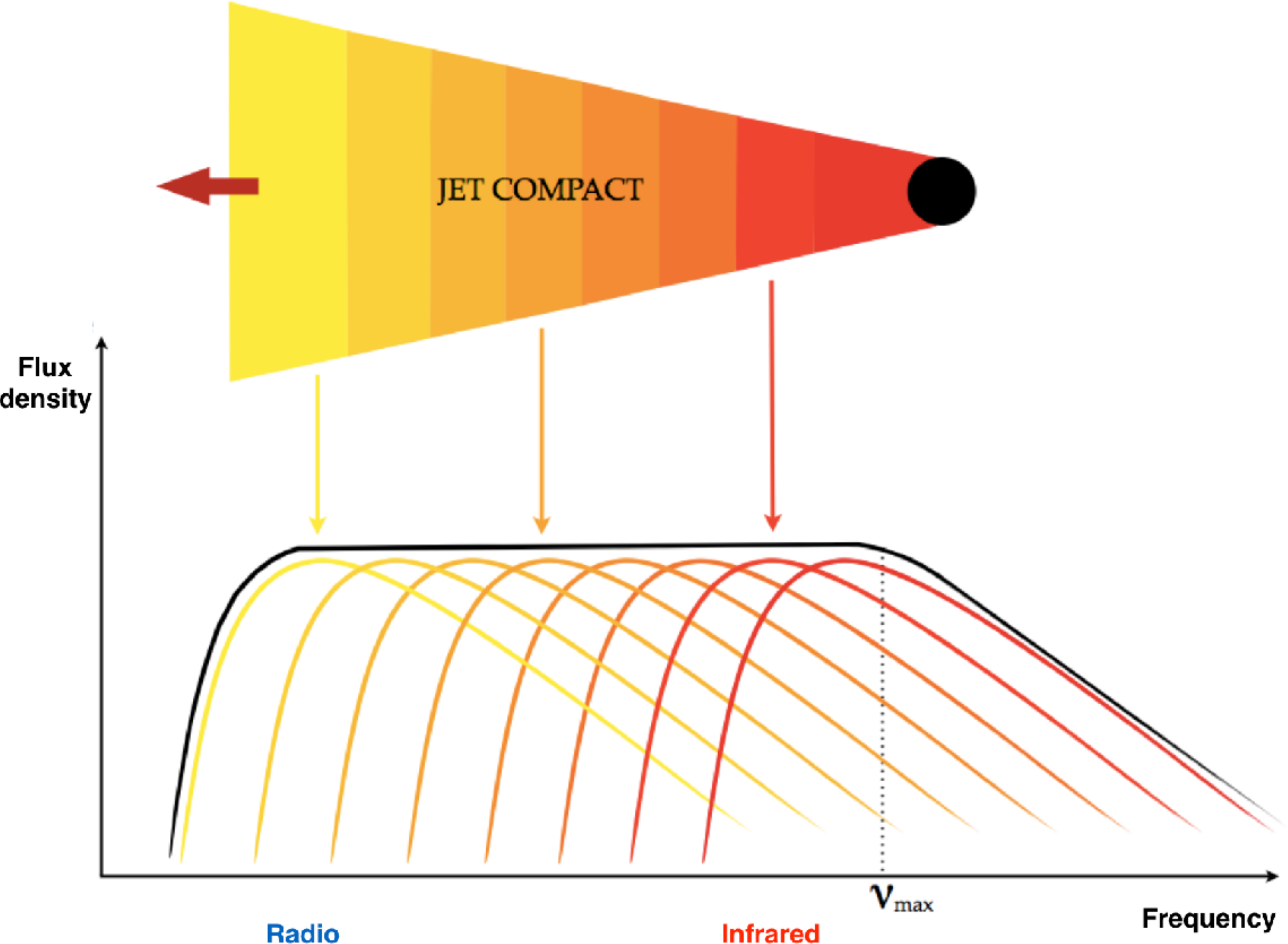}
\end{center}
\caption{
The spectral energy distribution of the jet which is made up of different 
synchrotron emitting regions along the length of the jet \cite{CoriatPHD10}.
}
\label{fig:sed-picture}       
\end{figure}

\subsection{The spectral energy distribution}

\begin{figure}
\begin{center}
  \includegraphics[scale=0.45]{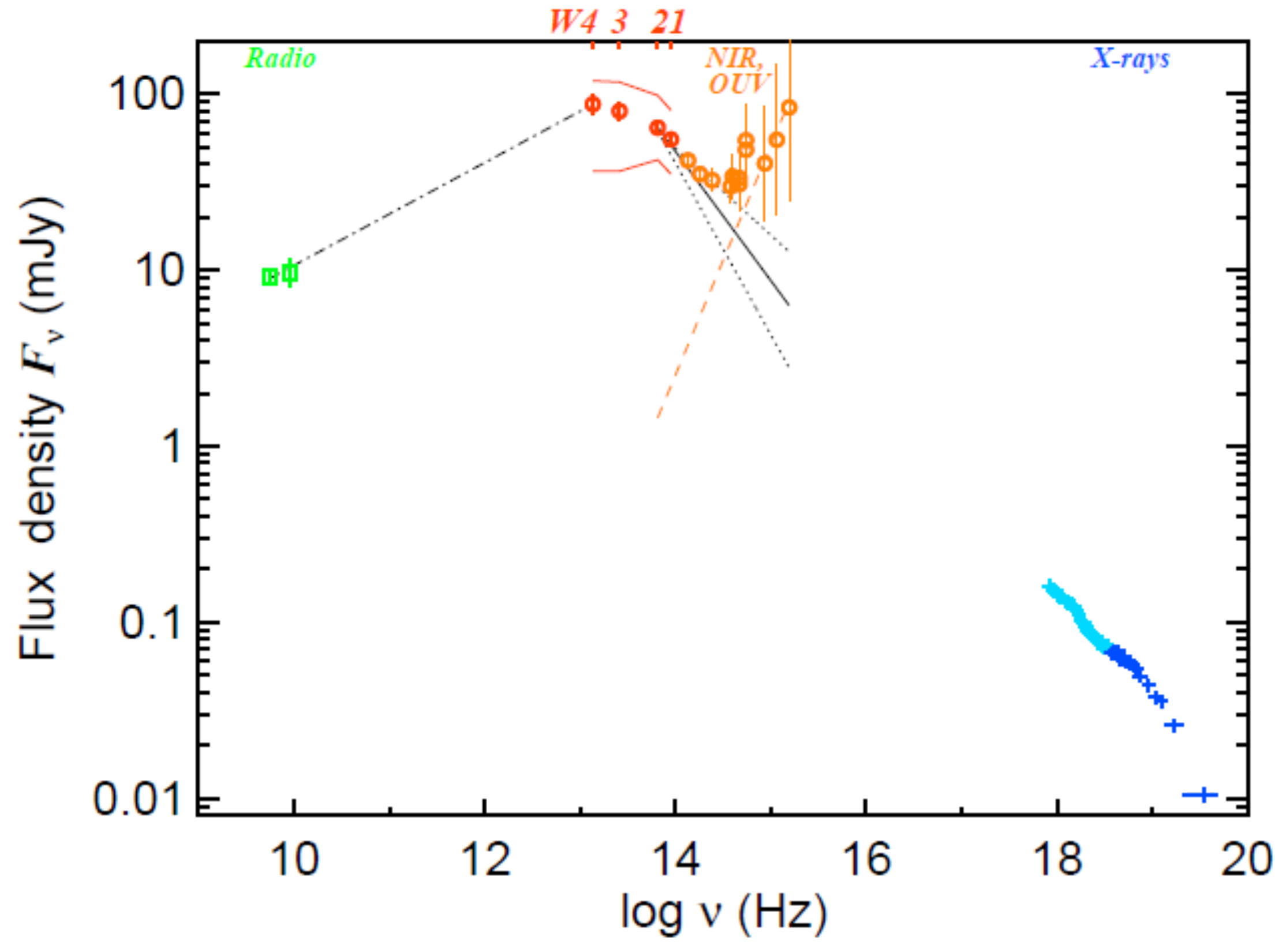}
\end{center}
\caption{
The spectral energy distribution of GX\,339--4. The spectrum breaks (jet break) from optically thick to optically thin in the mid-IR \cite{Gandhi11}. \copyright AAS. Reproduced with permission.
}
\label{fig:jet_break}       
\end{figure}

The spectral energy distribution of an XRB can be described by thermal 
emission from a companion star and accretion disc, synchrotron emission 
from a jet and a Comptonized corona from the disc. At optical and IR 
wavelengths the companion star dominates the optical/near-IR emission 
whereas the accretion disc dominates at UV to X-ray wavelengths. 
The radio to IR spectrum of an XRB in the hard-state is due to a $\sim$ 
flat, self-absorbed optically thick synchrotron spectrum with a spectral 
index of $\alpha \sim$ 0.0 to +0.5, where $F_\nu \propto \nu^\alpha $ 
\cite{Fender01, Fender00, Markoff01} (see Figs.\,\ref{fig:sed}, \ref{fig:sed-picture} 
and \ref{fig:jet_break}), which is the principal signature of the presence of 
a conical, collimated jet in black BH-XRBs. It typically extends at down 
to at least to millimeter wavelengths and is explained as being due to the 
superposition of synchrotron-emitting lepton particle distributions at 
different radii from the compact object \cite{Blandford79, Kaiser06, Malzac18}. 
(see Fig. \ref{fig:sed-picture}). 
The  higher energy synchrotron emission arises from a small, dense region of the 
jet, close to the location where the jets are launched, near the compact 
object \cite{Kaiser05}.  Since $\alpha >\sim$ 0, the bulk of the radiative 
power of the jet resides in this higher energy emission and so at some 
frequency ($\sim$ IR wavelengths) there is 
a break in the jet spectrum (e.g. \cite{Corbel02,Russell06}).
Above this break frequency, the flat spectral 
component breaks to an optically thin spectrum corresponding to the point 
at which the entire jet becomes transparent, close to the jet base in the 
post-acceleration plasma
(see Fig. \ref{fig:sed-picture}). In addition, there is a cut-off in the jet 
spectrum at higher energies, in the X-ray region \cite{Markoff01,Maitra09} 
(see Fig. \ref{fig:sed}).

The transition between the optically thick and thin part of the spectrum 
occurs at the jet break frequency, which usually lies in the mid-IR to 
near-IR , where the optically thin synchrotron emission is observed, 
resulting in a power law with index -1 $< \alpha <$ -0.5 \cite{Gandhi11, Russell13a} 
(see Fig. \ref{fig:jet_break}). The break frequency has been detected in a 
the BH candidates
GX\,339--4 \cite{Corbel02, Gandhi11, Corbel13},  
XTE\,J1118+480 \cite{Hynes06}, XTE\,J1550--564 \cite{Chaty11},  
V404\,Cyg \cite{Russell13b, Tetarenko19}, 
MAXI\,J1836--194 \cite{Russell13a, RussellT14a}, 
Cyg\,X--2 \cite{Rahoui11},
in the quiescenct BH-XRBs, Swift\,J1357.2--933 and A0620--00 
\cite{Plotkin16, Dincer18, Russell18}, and in the NS systems 4U\,0614+091, 
4U\,1728--34, and Aql\,X-1 \cite{Migliari10, DiazTrigo17, DiazTrigo18}. The 
steady jet spectrum evolves with the break frequency (from optically thick 
to optically thin) moving as a function of time and spectral state as the 
electron population cools \cite{RussellT14a}. During an XRB outburst, as the 
X-ray spectrum of the source softens, the jet break frequency moves from 
the near-IR to radio frequencies and then return to the near-IR at the end 
of the outburst, when the spectrum gets harder \cite{Russell13a, 
DiazTrigo18}.  In BH-XRBs, as the source softens during state transitions, 
the jet break is observed to shift from the IR to millimetre frequencies 
\cite{Koljonen15}.  In the soft X-ray state, no core jet is detected. The 
structure of the accretion flow is governed by the presence of large-scale 
height magnetic fields that determine the position of the break frequency. 
Since the size of the jet-emitting region scales 
inversely with the frequency \cite{Blandford79} the break frequency 
gives information on the size of the base of the jet where the particles 
are accelerated \cite{Chaty11, Ceccobello18}.

Fast sub-second optical/near-IR variability and correlations have also been 
observed in various systems \cite{Durant08, Durant11, Gandhi08, Gandhi10}. 
The most likely explanations is a jet origin, because the variability has a 
spectrum consistent with optically thin synchrotron emission \cite{Hynes03, 
Hynes06}, which is stronger in the near-IR \cite{Casella10, Gandhi10}, 
maybe due to variable synchrotron emission of the X-ray corona 
\cite{Veledina11}.

\begin{figure}[t]
\begin{center}
  \includegraphics[scale=0.5]{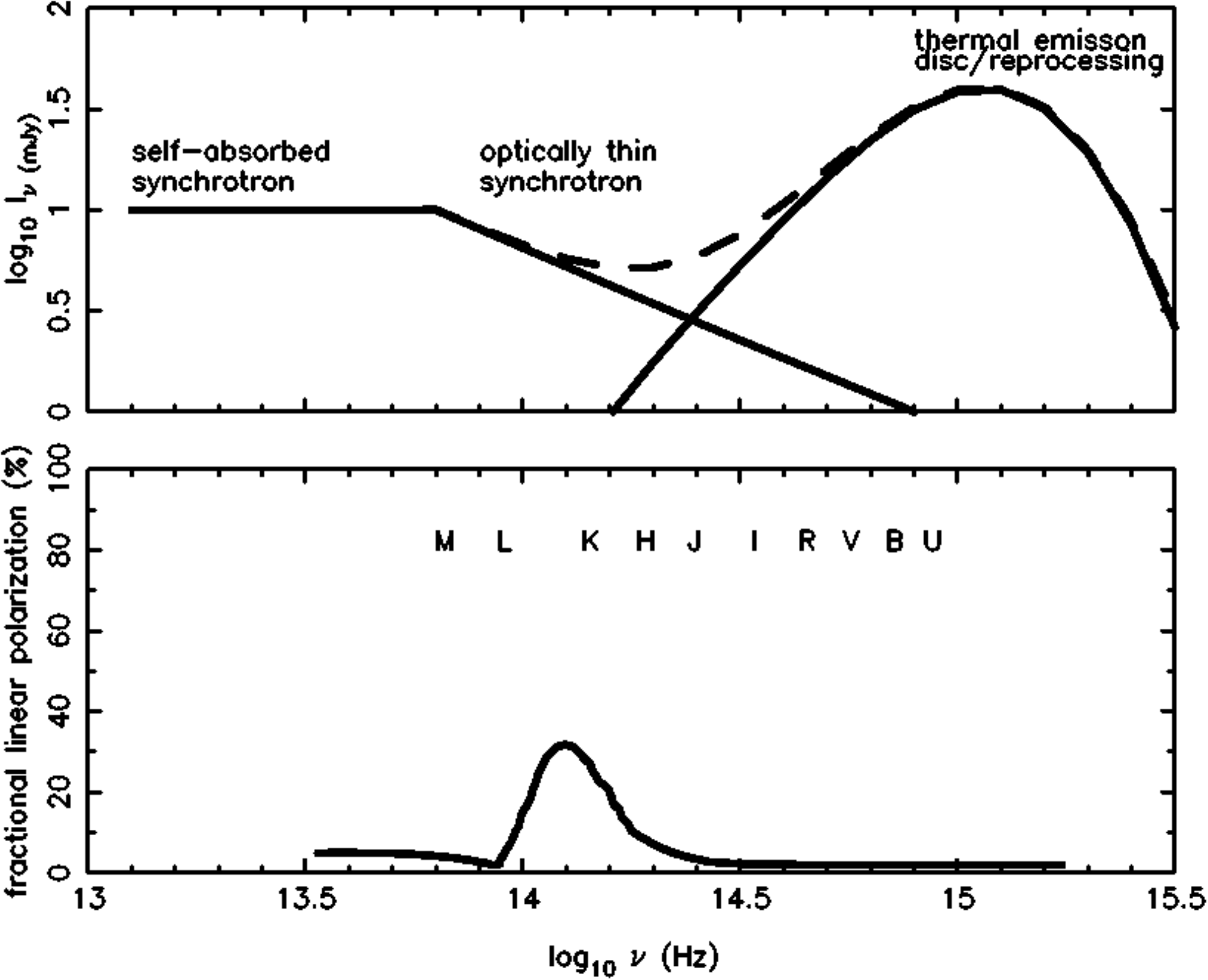}
\end{center}
\caption{
The expected linear polarization signature in the optical and near-IR
spectrum of XRBs. The key spectral points are  the break from optically thick (low
polarization $<$10\%) to optically thin (high polarization $<$70\%)
synchrotron emission, and the point at which the thermal (low-polarization)
emission begins to dominate over the jet. These effects should combine to
produce a narrow spectral region with relatively high linear polarization
arising from close to the base of the jet \cite{Shahbaz08}.
}
\label{fig:pol_sed}       
\end{figure}

\subsection{The polarization spectrum of XRBs}

Shahbaz et al. (2008) outlined the expected intrinsic linear polarization 
signature of XRBs in the optical/near-IR region. The optical and IR light 
from XRBs is polarized due to intrinsic polarization of the emitting 
photons, or by the (Thompson or possibly Rayleigh) scattering or absorption 
of unpolarized photons.  Hydrogen in the disc is totally ionised and a low 
linear polarisation is expected in the optical due to Thomson scattering of 
emitted unpolarised radiation with free electrons in the disc. In the case 
of the persistent XRBs where the accretion disc dominates the optical 
light, the linear polarization has a component that is produced by electron 
scattering by plasma above the accretion disc. Although a strong 
interstellar component is present, for Rayleigh and Thompson scattering 
processes, the level of linear polarization decreases as a function of 
increasing wavelength. Electron scattering of the disc radiation by the 
secondary star is insignificant, because the secondary stars in XRBs have 
low surface temperatures and therefore low free electron densities. 
Polarization of the optical light due to absorption by interstellar dust is 
inferred by the characteristic constant wavelength-dependent polarization, 
given by Serkowski's law \cite{Serkowski73, Schultz04}. There also exists 
the possibility of a large fractional linear polarization (FLP) from optically 
thin synchrotron emission. Indeed observations and theoretical results 
suggest a break between optically thick and thin emission occurs around the 
near-IR band (e.g., see \cite{Corbel02, Russell13b} and Fig. 
\ref{fig:jet_break}).

Optically thick synchrotron emission has a maximum linear polarization of 
$\sim$10\%, whereas optically thin synchrotron emission can have a FLP as 
high as 70\% \cite{Blandford02}. Two polarization signatures are expected 
from the jet. For the optically thick part of the spectrum, which results 
in the flat spectral component observed in the hard X-ray state 
\cite{Fender01}, no more than a few \% polarization is expected.  At long 
wavelengths there should be $\sim$1\% polarization from the self-absorbed 
jet and at short wavelengths a comparable level will be measured due to 
scattering in the accretion flow. However, in the relatively narrow 
spectral region in which optically thin synchrotron emission dominates, one 
expects a strong signature that initially rises to longer wavelengths as 
the relative jet/disc fraction increases (see Fig. \ref{fig:pol_sed}).

\begin{figure}[htbt]
\begin{center}
  \includegraphics[scale=0.3]{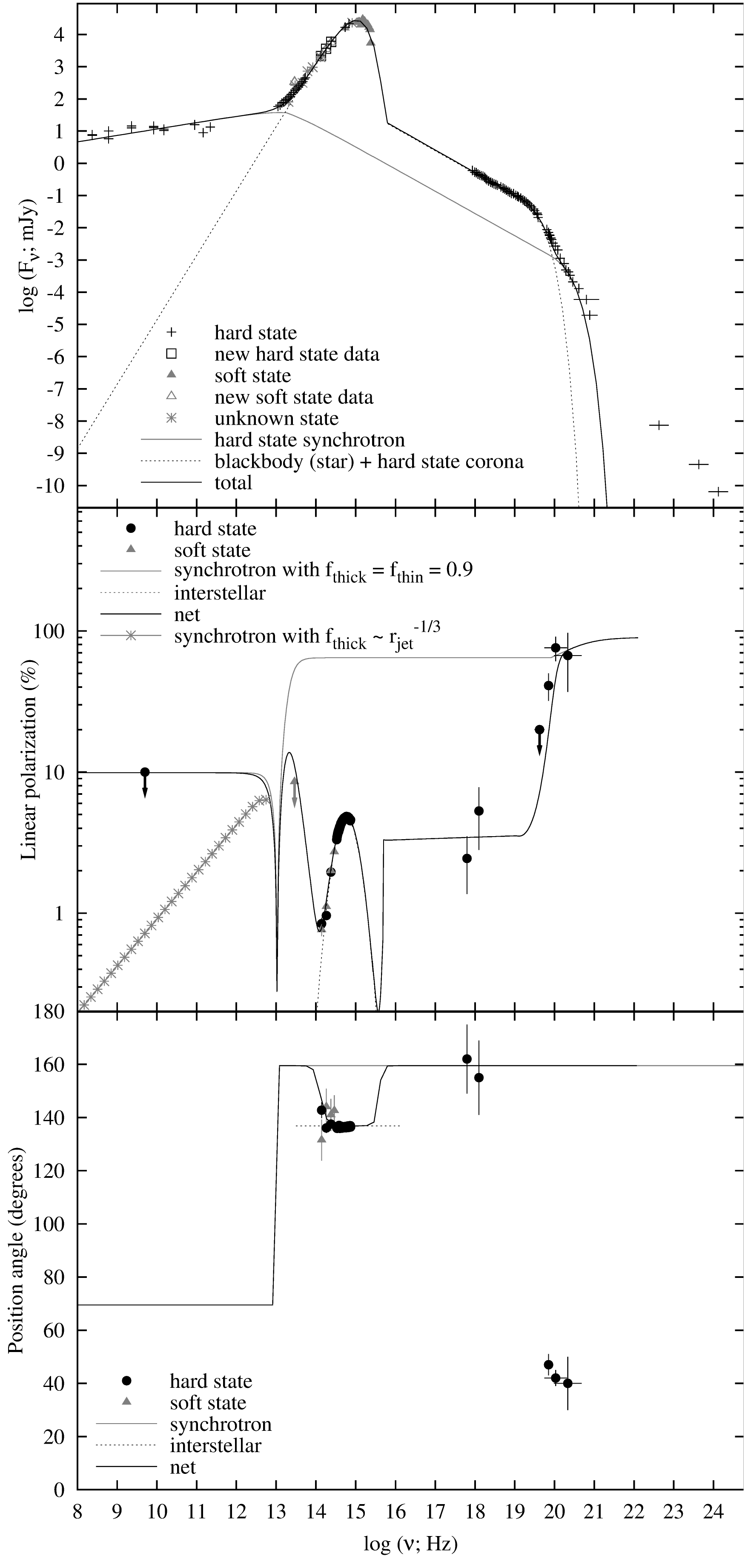}
  \hspace{2mm}
  \includegraphics[scale=0.3]{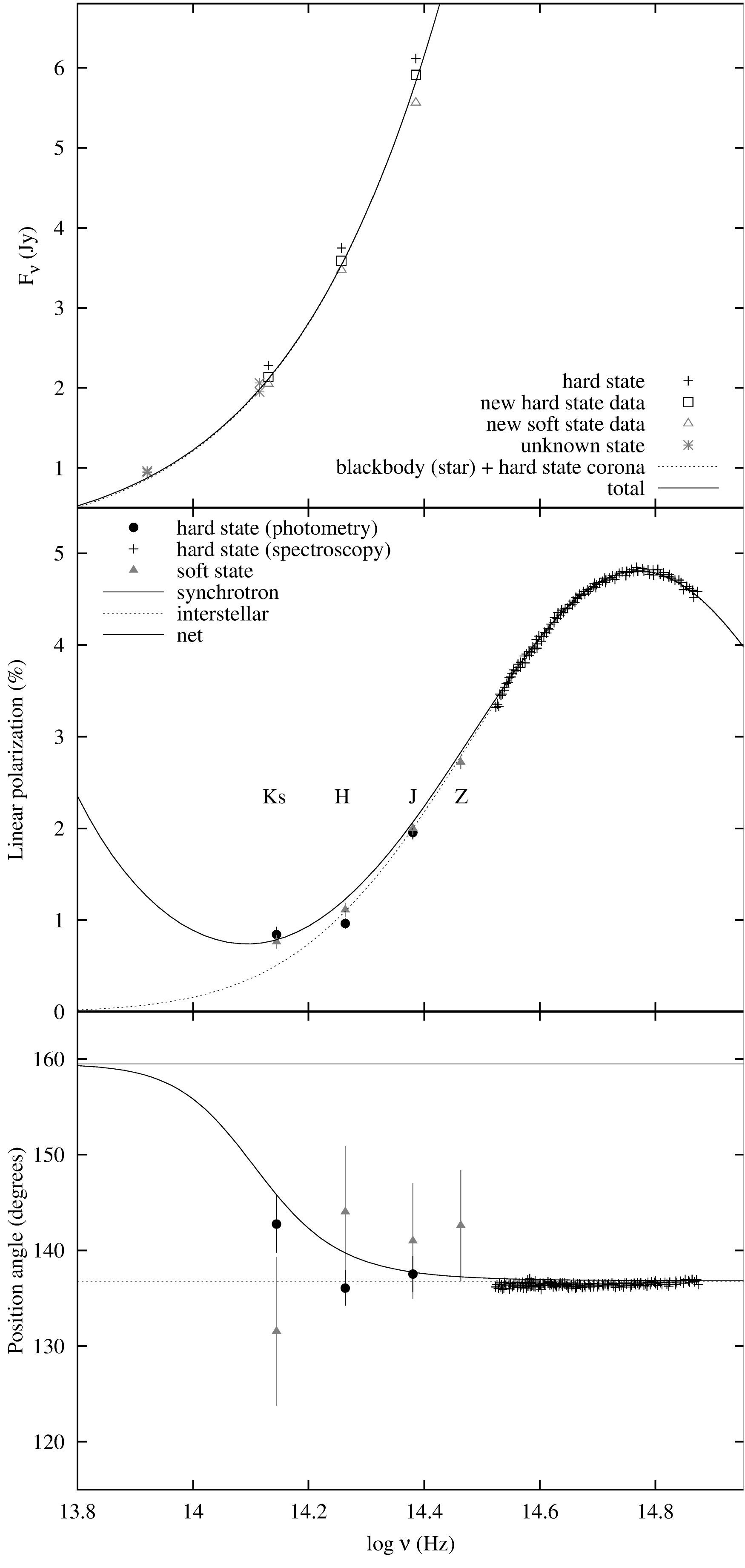}
\end{center}
\caption{Left: Radio to $\gamma$-ray flux density  (top), linear polarization 
spectrum (middle) and polarization position angle (bottom) spectrum of Cyg\,X--1 
Right: the same as the left-hand panels but just the near-IR/optical region \cite{Russell14b}.
 }
\label{fig:spec-pol-cygx1}       
\end{figure}

\subsection{The theoretical polarization spectrum of the jet}
\label{sec:pol_jet}

Relativistic electrons moving in a large-scale and ordered, uniform magnetic field 
produce synchrotron radiation. If the relativistic electrons produced by 
the first-order Fermi acceleration have a non-thermal energy distribution, 
the synchrotron radiation is linearly polarized. The linear 
polarization spectrum of optically thin synchrotron emission from an 
relativistic electrons in a uniform magnetic field was first treated by 
Westfold (1959). Subsequent works have extended these ideas to include 
power-law electron energy distributions and non-uniform magnetic fields 
(e.g. \cite{Ginzburg65,Nordsieck76,Longair11}). Indeed, Bjornsson et al. 
(1982) derived the more general case for optically thin synchrotron 
emission from an electron population and magnetic field with an arbitrary 
distribution of energies and configuration, respectively.  They showed that 
the expected FLP is given by

\begin{equation}
\rm FLP_{\rm thin} = f \frac{p+1}{p+7/3}  
               = f \frac{1-\alpha_{\rm thin}}{5/3-\alpha_{\rm thin}} 
\label{eqn:sed_1}
\end{equation}

\noindent
where $\alpha_{\rm thin}$ [=(p-1)/2] is the spectral index of the optically 
thin synchrotron emission of the form $\rm F_{\nu} \propto\nu^\alpha$, $f$ 
represents the ordering of the magnetic field and $p$ is the electron 
energy distribution ($\rm N(E)dE \propto E^{-p} dE$)  \cite{Rybicki86}. 
The parameter $f$ takes values between 0 and 1, where 0 represents non-uniform, no 
net field orientation and 1 a perfectly uniform, fully aligned field.  If 
the field is tangled, the differing angles of the polarized light suppress 
the observed, average polarization.  In the case of a fully ordered field 
\cite{Rybicki86, Bjornsson82} the maximum polarization strength is 
$\sim$70--80\% and is dependent only on the degree of ordering of the field 
and the energy distribution of the electron population. However, if the 
spectral index is steeper the polarization can be higher. The electric 
(polarization) vector is perpendicular to the projection of the magnetic 
field on the plane of the sky.

For a synchrotron source with a power-law electron energy distribution the 
FLP is frequency-independent. However, a high energy cutoff in the electron 
distribution results in frequency-dependent FLP \cite{Bjornsson85}.  Above 
the high energy cutoff frequency the polarization is given by

\begin{equation}
\rm FLP_{\rm cut}   = f \frac{1-\alpha_{\rm cut}}{5/3-\alpha_{\rm cut}} 
\label{eqn:sed_2}
\end{equation}

\noindent
where $\alpha_{\rm cut}$ is the spectral index defined by power-law 
emission model. The above situations corresponds to a highly idealized 
situation, assuming optically thin plasma, isotropic distributions of 
electrons, perfectly ordered homogeneous magnetic fields, and the absence 
of substantial perturbations. A major modification occurs for optically 
thick plasma where each light ray experiences multiple scattering events. 
For optically thick, absorbed synchrotron emission ($F_{\nu} \propto 
\nu^{5/2}$) the net polarization spectrum is given by

\begin{equation}
\rm FLP_{\rm thick} = f \frac{3}{6p+13}  
\label{eqn:sed_3}
\end{equation}

\noindent
with a polarization position angle (PA) that differs by 90 degrees to that 
of optically thin synchrotron polarization \cite{Pacholczyk70, Rudnick78, 
Blandford02,Longair11}. The polarization is less polarized than optically 
thin synchrotron \cite{Blandford02}. For optically thin synchrotron 
emission with a typical spectral index of $\alpha_{\rm thin}=-0.7$ ($p$ = 
2.4), FLP = 11\%.  The electric vector in this case is parallel to the 
projected magnetic field. If the level of ordering of the magnetic field is 
constant along the jet, with a constant PA, then Equation \ref{eqn:sed_3} 
can be used to describe the polarization expected from a flat/inverted 
optically thick jet spectrum.

An outlined in detail in Russell \& Shahbaz (2014), equations 
\ref{eqn:sed_1} to \ref{eqn:sed_3} can used to predict the FLP as well as 
the polarization PA of the synchrotron emitting jet spectrum. Given that 
different components produce emission with different FLPs and PAs, it is 
necessary to calculate the Stokes parameters $Q$ and $U$ for each component 
at each frequency from the known FLP and PA values, using the standard 
equations relating polarization and PA.

\subsection{Polarization in X-ray persistent XRBs}
\label{sec:pol_xrbs}

At optical/near-IR wavelengths, components such as the accretion disc and 
companion star often dominate the spectrum of BH- and NS-XRBs, suppressing 
any synchrotron contribution to the polarization. However, when the 
synchrotron makes a strong contribution, intrinsic polarization is detected 
\cite{Dubus06, Shahbaz08, Russell08, Russell11, Chaty11, Baglio14}.  
Intrinsic near-IR polarization has been detected in three BH-XRBs during 
outburst, GRO\,J1655--40, XTE\,J1550--564 and GX\,339--4, all during the 
hard-state when jets are expected to be launched \cite{Russell08, Chaty11, 
Russell11}.

\begin{figure}[htbt]
\begin{center}
  \includegraphics[scale=0.5,angle=-90]{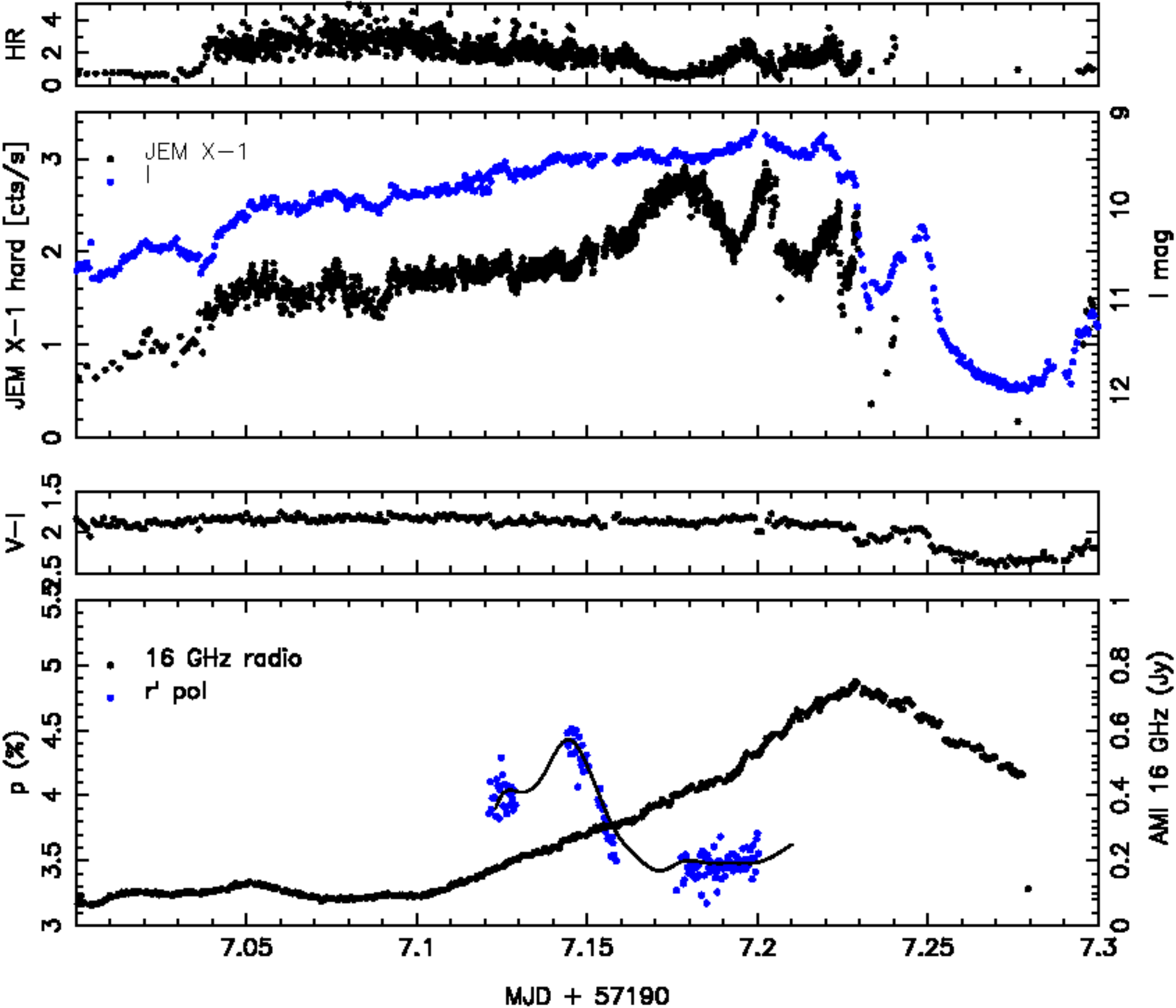}
\end{center}
\caption{
The X-ray, optical and radio light curves of the V404\,Cyg during the 
outburst in 2015. around MJD\,57197. From top to bottom: INTEGRAL JEM-X 
X-ray hardness ratio;  INTEGRAL JEM-X X-ray count rate (black dots) with the 
AAVSO $I$-band light curve (blue dots); AAVSO optical colour; and the AMI-LA 
16 GHz radio data with the $r'$-band polarization light curve (blue dots) 
\cite{Shahbaz16}.
}
\label{fig:v404}       
\end{figure}

\begin{figure}[b]
\begin{center}
  \includegraphics[scale=0.2]{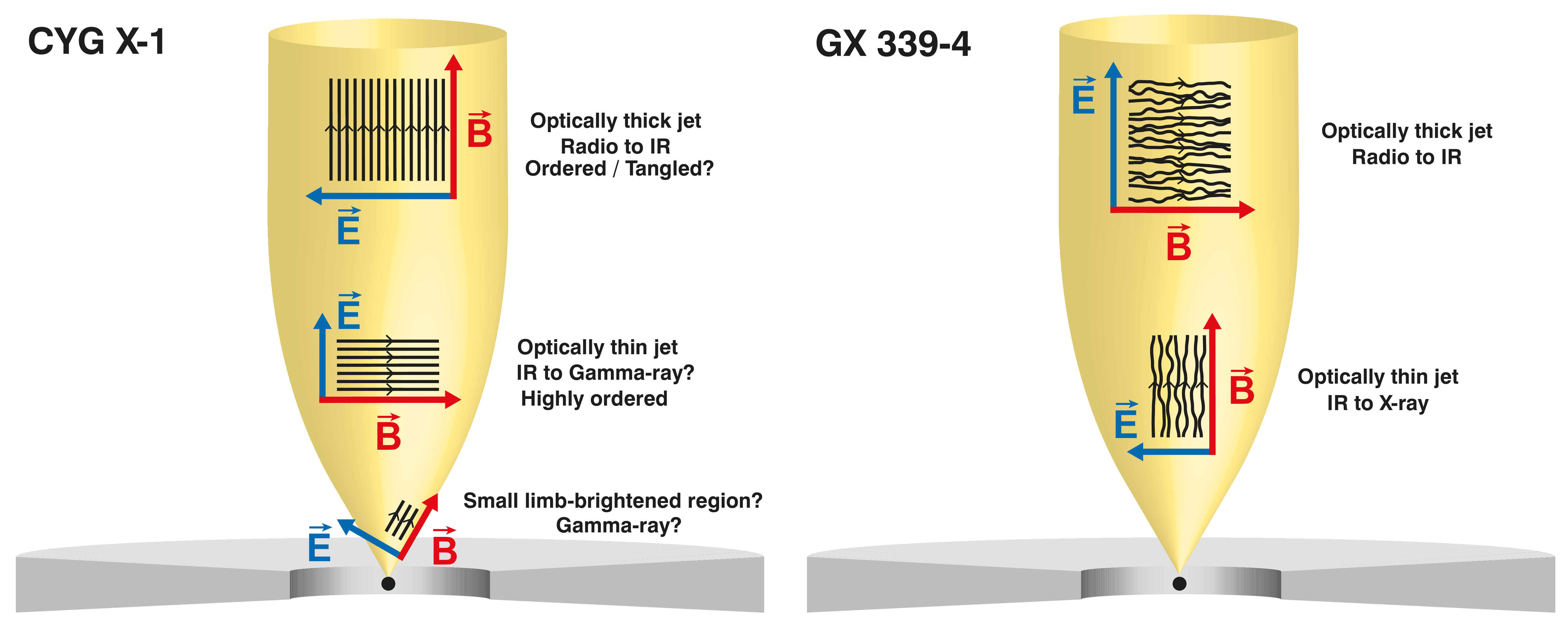}
\end{center}
\caption{
A schematic diagram of the Cyg\,X--1 jet showing the orientation of the 
electric (from the near-IR observations) and magnetic field vectors (from 
observations of the radio jet) \cite{Russell14b}. For synchrotron, the 
polarization PA traces the electric field vector in the 
emitting region, and the magnetic field vector differs by 90 degrees. In 
Cyg\,X-1 (left) a highly ordered magnetic field in the optically thin 
region near the jet base (as implied by the observations) and an ordered or 
tangled field in the large scale jet is shown.  In other XRBs, such as 
GX\,339--4, (right), the magnetic fields near the base of BH-XRB jets are 
moderately tangled, with a preferential orientation of the magnetic field 
along the jet axis.
}
\label{fig:Bfield_xrbs}       
\end{figure}

In GX\,339--4, when the polarization is strong the polarization angle is 
perpendicular to the known jet axis, implying a magnetic field that is 
parallel to the jet axis (see Fig. \ref{fig:Bfield_xrbs}). The low levels 
of polarization measured ($\sim$1--7 \%) and variability imply a tangled, 
rapidly changing magnetic field geometry near the base of the jet 
\cite{Russell08, Russell11}, Mid-IR variability of GX\,339--4, also show 
that either the acceleration region near the base of the jet, or its 
magnetic field strength, are changing on timescales of minutes 
\cite{Gandhi11}, the same timescale as the polarization variability. During 
the bright outburst of V404\,Cyg in 2015, flares of optical polarization 
were also detected and interpreted as synchrotron emission from a variable 
jet \cite{Lipunov16, Shahbaz16}.  Time-resolved optical polarimetry 
revealed a polarization flare which decayed fairly smoothly from 4.5 to 3.5 
\% 20 min, with a stable PA that is aligned with the resolved radio jet. 
This implies that the electric field vector near the base of the jet in 
V404\,Cyg is on average approximately parallel to the jet axis and 
that the magnetic field is orthogonal to the jet axis. This may be due to 
the compression of magnetic field lines in shocks in the flow, resulting in 
a partially ordered transverse field. The polarization flare also occurs 
during the initial rise of a major radio flare event that peaks later, and 
is consistent with a classically evolving synchrotron flare from an 
ejection event (see Fig. \ref{fig:v404}).

To date, the most detailed polarization spectrum has been obtained for 
Cyg\,X--1. Russell1 \& Shahbaz (2014) modelled the broadband, radio to 
$\gamma$-ray flux spectrum and polarization spectrum of Cyg\,X--1 in the 
hard-state, with a simple phenomenological model consisting of a strongly 
polarized synchrotron jet (see Sect. \ref{sec:pol_jet}) and an unpolarized 
Comptonized corona and a moderately polarized interstellar dust component 
(see Fig. \ref{fig:spec-pol-cygx1}).  They were able to show that the 
origin of the $\gamma$-ray, X-ray and some of the IR polarization is the 
optically thin synchrotron power law from the inner regions of the jet.

In NS-XRBs, intrinsic near-IR or optical polarization at similar levels as 
the BH-XRBs has detected in Sco\,X--1, Cyg\,X--2 and 4U\,0614+09 
\cite{Russell08, Shahbaz08, Russell11, Baglio14}. Shahbaz et al. (2008) 
performed near-IR spectropolarimetry of the XRBs Sco\,X--1 and Cyg\,X--2 
and found them be intrinsically polarized, with an increasing FLP at lower 
frequencies, which is explained by optically thin synchrotron emission from 
a jet. The FLP observed is $\sim$1--10\%, with evidence for rapid 
variations in some sources on timescales of seconds--minutes 
\cite{Russell11}. This dependence on wavelength cannot be explained by 
scattering or interstellar dust absorption and it cannot explain the short 
timescale variability of the polarization observed Sco\,X--1 
\cite{Russell11}.  The PA is usually approximately orthogonal to the axis 
of the resolved radio jet, which implies the magnetic field is parallel to 
the jet axis. In general, the observations of most XRBs are consistent with 
a variable, predominantly tangled magnetic field geometry, with field lines 
preferentially orientated along the jet axis (see Fig. 
\ref{fig:Bfield_xrbs}). However, there are a few exceptions. Cyg\,X--1, has 
a highly ordered, stable magnetic field near the base of the jet that is 
perpendicular to the jet axis \cite{Russell14b, Rodriguez15}. V404\,Cyg 
shows a low level polarization flare, with the magnetic field also 
perpendicular to the jet axis \cite{Shahbaz16}, which may be due to the 
compression of tangled magnetic field lines in shocks downstream in the 
flow or collisions with dense regions of the interstellar medium, resulting 
in a partially ordered transverse field.

\subsection{Polarization in X-ray quiescent XRBs}

X-ray transients are a subclass of the XRBs which spend the majority of 
their time in a state of quiescence, accreting at very low X-ray 
luminosities $\rm \sim 10^{30}-10^{34}\,\,erg\,s^{-1}$, explained by a 
truncated accretion disc with a radiatively inefficient accretion flow 
\cite{Narayan94}. Although it is not really clear if quiescence is a separate state, 
or simply a low-luminosity version of the hard-state, but observations show 
that quiescent BH-XRBs have softer X-ray spectra than hard-state systems 
(e.g. \cite{Reynolds14}) and outflows with weaker particle acceleration 
\cite{Gallo07, Plotkin15}. There are a few radio detections of XRBs in 
quiescence, which include V404\,Cyg \cite{Gallo05}, A0620--00 
\cite{Gallo06}, XTE\,J1118+480 \cite{Gallo14}, Swift\,J1753.5-0127 
\cite{Miller-Jones11}, 47\,Tuc X9 \cite{Miller-Jones15}, M62-VLA1 
\cite{Chomiuk13}, MWC 656 \cite{Dzib15} and M15\,S2 \cite{Tetarenko16}, 
however, there is only one linear polarization constraint of the quiescent 
radio emission; $<$2.11\% in V404\,Cyg \cite{Rana16}.  
A0620--00 and V404\,Cyg are the only truly quiescent system with a robust 
radio detection that seem to follow the same radio/X-ray correlations seen 
in the hard-state \cite{Gallo06}.

\begin{figure}
\begin{center}
  \includegraphics[scale=0.45]{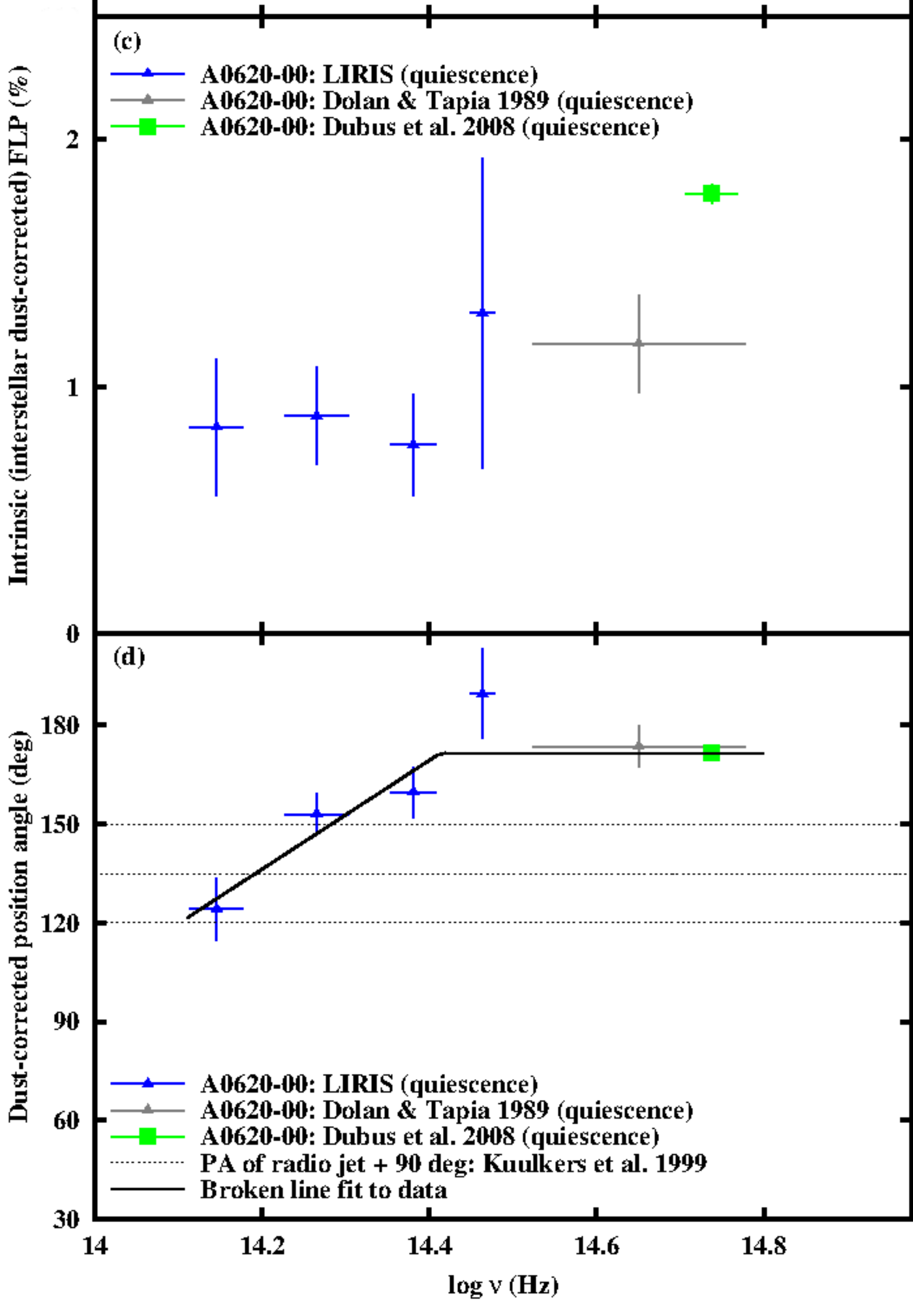}
\end{center}
\caption{
The intrinsic (dust-corrected) near-IR linear polarization and position 
angle spectra of A0620--00 in quiescence \cite{Russell16}. The 
dust-corrected FLP of A0620--00 is $ \sim$1--2\% in the optical and 
$\sim$0.5--1.0\% at near-IR wavelengths. The optical FLP is most likely due 
to Thomson scattering within the binary system \cite{Brown78}, 
which is observed to vary with orbital phase 
\cite{Dolan89b}. Note the different PA in the $H$- and $K_s$-bands compared 
to in the  optical. 
}
\label{fig:pol_a0620}       
\end{figure}

Jets in quiescent XRBs are weaker than the brighter hard-state jets. The 
electron may be dominated by partially non-relativistic, perhaps Maxwellian 
energy distributions, leading to steeper spectral energy distributions than 
are typical for optically thin synchrotron emission \cite{Shahbaz13, 
Markoff15}. Evidence for synchrotron emission from the optical/near-IR 
spectral energy distribution in quiescence has been found in the BH-XRBs 
A0620--00 \cite{Gallo07, Gelino10, Froning11, Dincer18}, 
Swift\,J1357.2--0933 \cite{Shahbaz13, Plotkin16} and the NS-XRB 
XTE\,J1814--338 \cite{Baglio13}. Near-IR polarimetric observations of two 
quiescent BH-XRBs has revealed the polarimetric signature of synchrotron 
emission \cite{Russell16}. In 
A0620--00, at the longest near-IR bands the PA of the excess 1.3\% 
polarization changes significantly (see Fig. \ref{fig:pol_a0620}). The PA 
of this weak synchrotron component, most likely from a jet, implies that 
the orientation of the magnetic field is approximately parallel to the 
radio jet axis \cite{Kuulkers99}. In Swift\,J1357.2--0933 the near-IR flux variability 
imply a continuously launched, highly variable jet, which has a moderately 
tangled magnetic field close to the base of the jet.  These results imply 
that weak jets in low luminosity accreting systems have magnetic fields 
which possess similarly orientations and tangled fields compared to the more luminous, XRBs at higher accretion rates (see Sect. \ref{sec:pol_xrbs}).

\subsection{Time-resolved polarimetry}
\label{sec:time}

Phase-resolved linear polarimetry provides a unique probe of XRBs as one 
can obtain information on geometrical properties of the binary system from 
the light scattered by circumstellar material. 
In principle the orbital inclination ($i$) can be measured, which 
is essential in determining the 
components' masses, the fundamental parameters that determines a star's 
initial structure and subsequent evolution. Brown, McLean \& Emslie (1978) 
and Rudy \& Kemp (1978) have investigated the polarization produced by Thomson 
scattering in circumstellar envelopes in binary systems.  They showed that 
the linear polarization has a characteristic variability around the binary 
orbit, associated with scattering in the stellar envelope that is distorted 
by the gravity of the compact object. One assume a circular binary orbit, 
scattering due to electrons and that the material is optically thin 
and co-rotates with the binary. The observed phase-resolved Stokes 
parameters $Q$ and $U$ light curves can be fitted with fit a truncated 
(first and second order) Fourier series,

\begin{align}
Q = q_0 + q_1\cos\lambda  + q_2\sin\lambda 
                  + q_3\cos2\lambda + q_4\sin2\lambda \nonumber\\
U = u_0 + u_1\cos\lambda  + u_2\sin\lambda 
                  + u_3\cos2\lambda + u_4\sin2\lambda 
\end{align}

\noindent 
where $\lambda=2\pi\phi$ is and $\phi$ is the orbital phase. The variations 
are usually dominated by the second harmonics, which is commonly seen for 
binary systems showing polarization modulation. In the $Q-U$ plane the 
orbital phase-resolved polarization traces an ellipse twice per orbit. One 
can derive the orbital parameters either by studying the geometry of the 
$Q-U$ plane or by considering the Fourier coefficients. A least-square fits 
can be performed on the observed $Q$ and $U$ light curves to determine the 
coefficients and $q_j$ and $u_j$ ($j$=0--4), which can then be compared 
with theoretical expectations

\begin{align}
Q = \tau_0\,[\,\,(1-3\gamma_0)\sin^2\,i \,+\, \sin\,2i(\gamma_1\cos\,\lambda 
\,-\, \gamma_2\sin\,\lambda) \nonumber\\
\,-\,(1+\cos^2\,i)(\gamma_3\cos\,2\lambda - \gamma_4\sin\,2\lambda)] \nonumber\\
U = 2\tau_0\,[\,\,\sin\,i(\gamma_1\sin\,\lambda \,+\, \gamma_2\cos\,\lambda) 
   \,-\, \cos\,i(\gamma_3\sin\,2\lambda + \gamma_4\cos\,2\lambda)]
 \end{align}

\noindent 
where $\gamma_i$ are moments of the scatterers's distribution 
\cite{Brown78,Drissen86}. The binary inclination is then given by

\begin{align}
\left[\frac{1-\cos\,i}{1+\cos\,i}\right]^2 = 
\frac{(u_1 + q_2)^2 + (u_2-q_1)^2}{(u_2 + q_1)^2 + (u_1-q_2)^2} \nonumber\\
\left[\frac{1-\cos\,i}{1+\cos\,i}\right]^4 = 
\frac{(u_3 + q_4)^2 + (u_4-q_3)^2}{(u_4 + q_3)^2 + (u_3-q_4)^2}
\end{align}

\noindent
Alternatively, in the $Q-U$ plane the orbital phase-resolved polarization 
traces an ellipse twice per orbit, and the eccentricity $e$ of this ellipse 
is related to the binary inclination by \cite{Rudy78}

\begin{align}
\cos i =  \sqrt \frac{1-e}{1+e}
\end{align}

\begin{figure}[t]
\begin{center}
  \includegraphics[height=4.5cm]{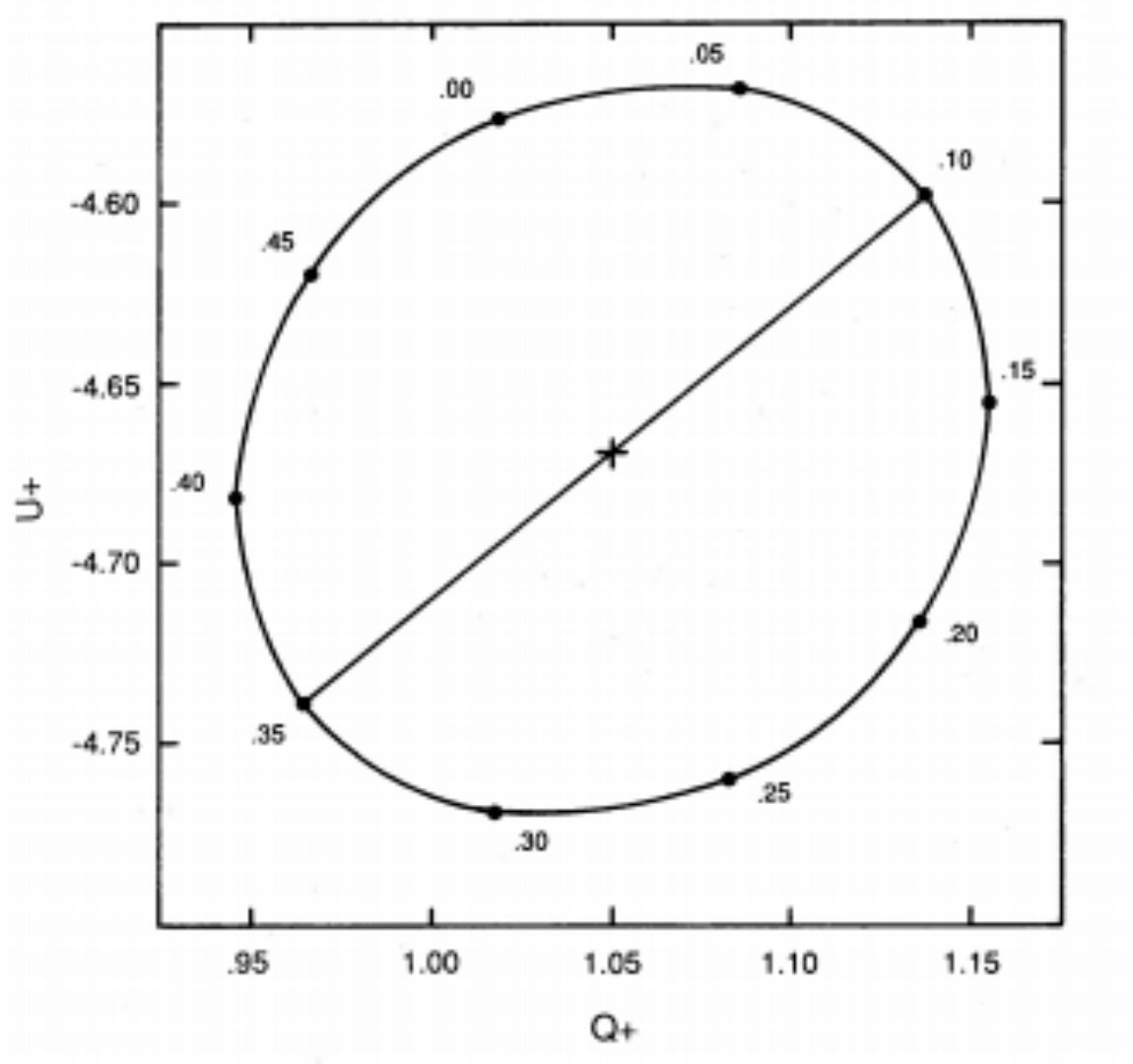}
  \includegraphics[height=4.5cm]{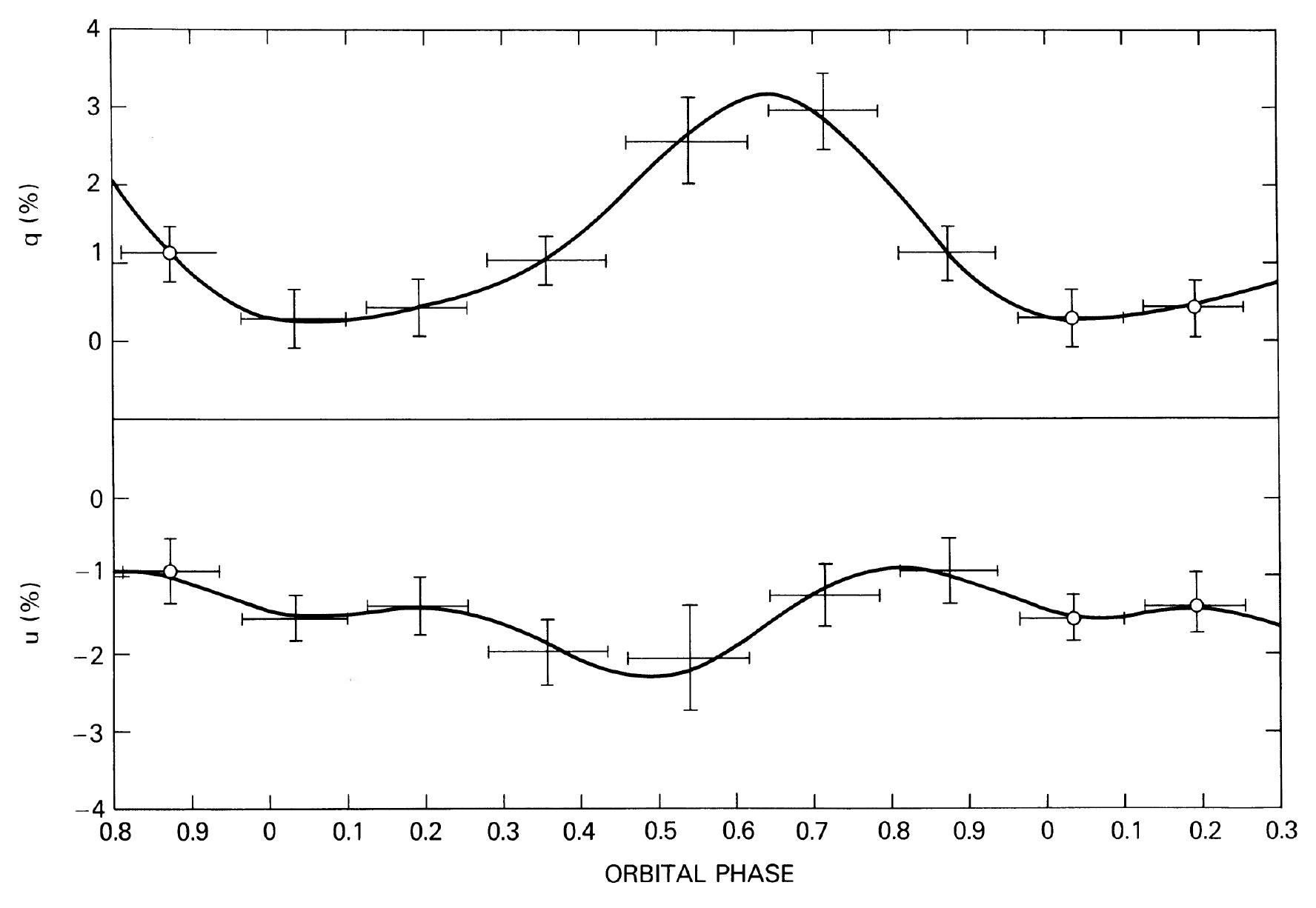}
\end{center}
\caption{ 
Left: The observed Stokes optical linear polarization of Cyg \,X--1 
in the $Q$-$U$ plane 
\cite{Dolan89a}. The values trace an ellipse, the eccentricity of which can be
used to determine the inclination angle.   Right: The observed $Q$ and $U$
time-resolved linear $V$-band polarimetric light curve of A0620--00
\cite{Dolan89b}. The solid line represents  the second order Fourier fit to the
data, which is then used to determine the inclination angle. 
\copyright AAS. Reproduced with permission.
} 
\label{fig:qu}       
\end{figure}

\noindent
Hence, by determining the eccentricity of observed $Q-U$ ellipse, one can 
determine the binary inclination, independent of size, shape and position 
of the circumstellar region (see Fig. \ref{fig:qu}). An extension of the 
above model to an elliptical binary orbit can by found in \cite{Brown82} 
and \cite{Simmons84}. Detailed accounts on polarimetric inclination 
determinations and their uncertainties are given by \cite{Aspin81, 
Simmons82, Wolinski94}.

Time-resolved optical polarimetric studies of two XRBs has allowed one 
to estimate the inclination angle; Cyg\,X--1 \cite{Kemp79,Dolan89a, 
Nagae09} and A0620--00 \cite{Dolan89b}.  In Cyg\,X--1 a $\sim$5\% optical 
linear polarization with an amplitude of $\sim$0.3\% is observed, whereas 
in A0620--00 a 3\% linear polarization with a $\sim$2\% modulation is 
observed on the orbital period (see Fig. \ref{fig:qu}). Using the method 
outlined above, the inclination angle been estimated to be $i$=62 (+5,-37) 
degrees and $i$=57 (+20,-40) degrees for Cyg\,X--1 and A0620--00, 
respectively \cite{Dolan89a,Dolan89b}. The uncertainties in the values 
large mainly because of the poor phase-sampling and precision of the Stokes 
light curves.

\begin{acknowledgement}
I would like to thank Dave Russell for the many discussions on the works 
presented in this review.
\end{acknowledgement}


\end{document}